\documentclass[a4paper,12pt,parskip=half,english,numbers=noenddot]{scrartcl}

\usepackage{babel}
\usepackage[T1]{fontenc}
\usepackage{lmodern}
\usepackage{verbatim}
\usepackage{textcomp,latexsym}
\usepackage{cancel}
\DeclareOldFontCommand{\it}{\normalfont\itshape}{\mathit}

\usepackage{amsthm,amssymb,amsmath}
\usepackage{bm}

\usepackage{paralist,threeparttable}
\usepackage[footnotesize,labelfont={bf}]{caption}
\usepackage{alphalph}
\usepackage{longtable,booktabs,multirow}
\usepackage{tabularx,nth}
\usepackage{enumerate}
\usepackage{ulem,upgreek}

\usepackage{tikz}
\usepackage{pifont}
\usepackage{graphicx,psfrag,color,rotate}
\graphicspath{{./pictures/}}

\usepackage{hyperref}
\usepackage[boxed]{algorithm2e}

\DeclareCaptionStyle{table_new}{font={footnotesize},justification=raggedright}
\setdefaultenum{1)}{\theenumi.1)}{}{}

\makeatletter
\renewcommand*{\env@matrix}[1][*\c@MaxMatrixCols c]{%
  \hskip -\arraycolsep
  \let\@ifnextchar\new@ifnextchar
  \array{#1}}
\makeatother

\newtheorem{remark}{Remark}

\newcommand{\indi}[1]{\ensuremath{\operatorname{#1}}}
\newcommand{\op}[1]{\ensuremath{\operatorname{#1}}}

\makeatletter
\DeclareRobustCommand\bigop[1]{%
  \mathop{\vphantom{\sum}\mathpalette\bigop@{#1}}\slimits@
}
\newcommand{\bigop@}[2]{%
  \vcenter{%
    \sbox\z@{$#1\sum$}%
    \hbox{\resizebox{\ifx#1\displaystyle.9\fi\dimexpr\ht\z@+\dp\z@}{!}{$\m@th#2$}}%
  }%
}
\makeatother

\DeclareMathOperator*{\arginf}{arginf}

\renewcommand{\vec}{\boldsymbol}

\renewcommand{\d}[1][]{\ensuremath{\,\mathrm{d}#1}}

\newcommand{\nablaq}[1]{\ensuremath{\nabla_{\hspace{-0.9mm}\mbox{\begin{scriptsize}$\mathrm{#1}$\end{scriptsize}}}}\hspace{0.3mm}}

\renewcommand{\d}{\operatorname{d}\!}

\newcommand{\RVE}{\ensuremath{\Omega_0}}

\newcommand{\dcross}{\mathrel{\vcenter{\offinterlineskip\ialign{##\cr\footnotesize{$\times$}\cr\noalign{\kern-1pt}\footnotesize{$\times$}\cr}}}}

\usepackage{tikz}
\usetikzlibrary{decorations.pathmorphing} 
\usetikzlibrary{matrix} 
\usetikzlibrary{arrows, arrows.meta} 
\usetikzlibrary{calc} 
\usetikzlibrary{shapes}
\usetikzlibrary{shapes.geometric}

\usepgfmodule{nonlineartransformations}
\usepgflibrary{curvilinear}

\usepackage{tikz-3dplot} 

\tikzstyle{block} = [draw,rectangle,thick,minimum height=2em,minimum width=2em]
\tikzstyle{sum} = [draw,circle,inner sep=0mm,minimum size=2mm]
\tikzstyle{connector} = [->,thick]
\tikzstyle{line} = [thick]
\tikzstyle{branch} = [circle,inner sep=0pt,minimum size=1mm,fill=black,draw=black]
\tikzstyle{guide} = []
\tikzstyle{snakeline} = [connector, decorate, decoration={pre length=0.2cm,
	post length=0.2cm, snake, amplitude=.4mm,
	segment length=2mm},thick, magenta, ->]

\begin{document}

\begin{center}
\large{\textbf{Variational formulation and monolithic solution of computational homogenization methods.}}

{\large Christian Hesch$^{a}$\footnote{Corresponding author. E-mail address: christian.hesch@uni-siegen.de}, Felix Schmidt$^{a}$,  Stefan Schu{\ss}$^{a}$}

{\small
\(^a\) Chair of Computational Mechanics, University of Siegen, Germany
}

\end{center}

\vspace*{-0.1cm}\textbf{Abstract}
In this contribution, we derive a consistent variational formulation for computational homogenization methods and show that traditional FE$^2$ and IGA$^2$ approaches are special discretization and solution techniques of this most general framework. This allows us to enhance dramatically the numerical analysis as well as the solution of the arising algebraic system. In particular, we expand the dimension of the continuous system, discretize the higher dimensional problem consistently and apply afterwards a discrete null-space matrix to remove the additional dimensions.  A benchmark problem, for which we can obtain an analytical solution, demonstrates the superiority of the chosen approach aiming to reduce the immense computational costs of traditional FE$^2$ and IGA$^2$ formulations to a fraction of the original requirements. Finally, we demonstrate a further reduction of the computational costs for the solution of general non-linear problems. 

\textbf{Keywords}: Variational, multiscale,  linearisation, FE$^2$, IGA$^2$, representative volume element, computational homogenization, null-space.
\section{Introduction}
Complex materials with dedicated morphological features like fibers or additive manufactured microstructures are nowadays used  throughout the industry. To obtain a suitable model,  continuum mechanical or thermomechanical formulations are often used based on infinitesimal strain measures.  Obviously, every possible finite inhomogeneity can be resolved by an infinitesimal model, however, the computational costs of such a large scale approximative solution may economically not affordable or in certain cases technically not possible. Additionally,  a material may appear perfectly homogeneous on the considered scale of approximations, whereas the microstructural information has been lost and often we are only interested in the macroscale solution.  Therefore, two different approaches are common to include this information. First, the microstructure can be taken into account in a phenomenological way.  The resulting constitutive relations can also incorporate size effects, leading to generalized theories for materials, see \cite{maugin_mechanics_2010,altenbach_mechanics_2011,bertram_mechanics_2020} for a general overview of gradient extended continua.  In contrast, microstructural information about morphology and material properties can be accounted for in a more explicit manner by means of homogenization methods, assuming that the scales of the finite approximation and the morphological features on the microscale are clearly separated. 

In the context of first-order homogenization schemes, we refer to \cite{hill_elastic_1952,hashin_variational_1963,hill_self-consistent_1965,mori_average_1973,willis_bounds_1977} for fundamental analytical approaches and to \cite{smit_prediction_1998,miehe_computational_1999-1,feyel_fe2_2000,kouznetsova_approach_2001} for seminal contributions to two-scale finite-element (FE) simulations.  For computational multiscale techniques using the well-known FE\textsuperscript{2}-methods we refer to the review article in  \cite{schroder_numerical_2014} and to \cite{matous_review_2017} for a general overview of computational multiscale techniques.  The application to solids at finite strains including physically coupled problems in thermo-elasticity are presented in \cite{TEMIZER2011344,TEMIZER201274} and for electro-elasticity in \cite{KEIP201462}.  The general extension on computational homogenization of strain gradient materials using Isogeometric Analysis, labeled IGA$^2$-method, is presented in our preliminary work in \cite{hesch2022}.  

Traditional methods like FE\textsuperscript{2} and  IGA$^2$ consider a representative volume element (RVE) at every Gauss point. Since the primary idea was to utilize the RVE and the corresponding set of partial differential equations representing the static equilibrium of some kind of a microstructure in exchange of a constitutive law. However,  we already know from concepts like the Hu-Washizu functional (see Zienkiewicz et al. \cite{zienkiewicz2005a, zienkiewicz1986}) that the relation between the strain energy function and the stress tensor can be considered as an independent equation in a weak form to be solved using again a finite element framework. To this end, the stress tensor is not evaluated at every Gauss point but rather interpolated using the applied shape functions for the solution of the weak form, see Bonet et al. \cite{bonet2015a} for details on the implementation of Hu-Washizu and Hellinger-Reissner type formulations.  

In this work, we propose a dimensional expansion of the macroscopic system for the calculations within the RVE. The arising microscopical fluctuations are interpolated in the expanded, higher-dimensional framework using continuous shape functions on the microscale and discontinuous shape functions on the macroscale to allow again for a static condensation procedure, written here in the form of a null-space reduction scheme.  With this monolithic framework, we can reconstruct traditional methods like FE\textsuperscript{2} and  IGA$^2$ by using Delta Dirac on the Gauss points and the application of a staggered scheme between the dimensions. However,  based on the variational formulation we can now make use of sophisticated methods for the solution of the arising large scale algebraic system of equations. As already shown in Lange et al.\ \cite{huetter2021}, monolithic solution techniques using static condensation procedures are by far superior to the original staggered scheme.  Both,  suitable choices of the macroscale shape functions and the application of sophisticated solution techniques lead to a dramatic reduction of the computational cost preserving the accuracy of the solution.  Additionally, this most general approach allows for the application of highly efficient methods like recursive trust-region multigrid formulations, see, among others, Gross \& Krause \cite{krause2009c}. 

The manuscript is organized as follows. First, the governing equations are presented in Section \ref{sec:basics} including the variational form of the multiscale problem.  The discretization and the arising algebraic system of the multiscale boundary value problem are shown in Section \ref{sec:multi}, followed by the different solution procedures in Section \ref{sec:solution}.  Representative examples are given in Section \ref{sec:numerics} and conclusions are drawn in Section \ref{sec:conclusions}.
\section{Governing equations}\label{sec:basics}
We start with a short summary of the governing equations for the macro- and afterwards for the microscale.  At the end of this section, we will introduce a common variational formulation for both scales. With regard to notation, Einstein's summation convention is used in the following for clearer presentation. 
\subsection{Macro-continuum}
\begin{figure}[htb]
\begin{center} 
\includegraphics[width=0.7\textwidth]{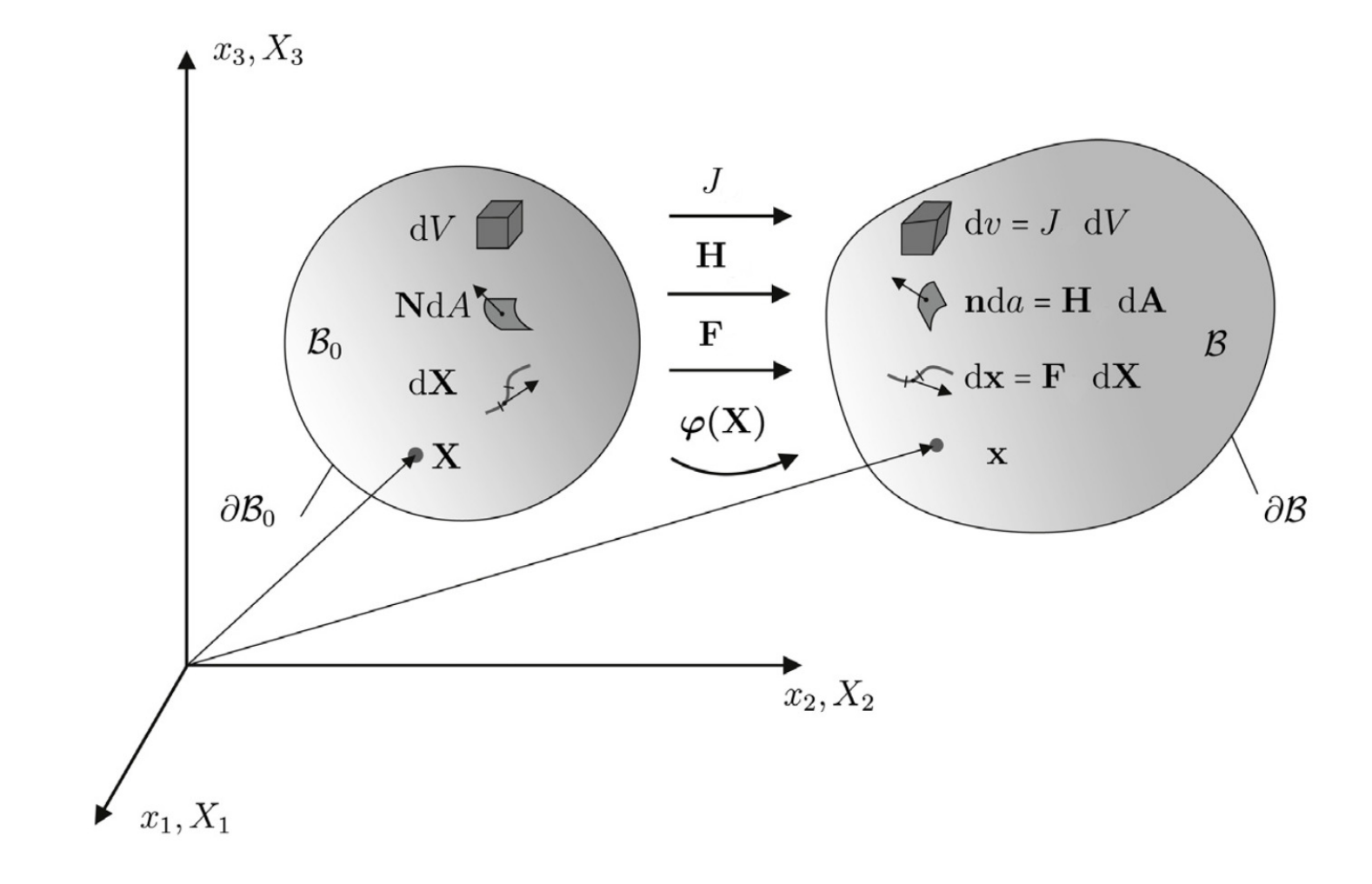} 
\caption{Graphical representation kinematical values $\vec{\varphi}$, $\vec{F}$, $\vec{H}$ and $J$, taken from \cite{BETSCH2018660}.}\label{fig:kin}
\end{center}
\end{figure}

Consider a Lipschitz bounded continuum body in its reference configuration \(\mathcal{B}_{0}\subset\mathbb{R}^{n}\) with \(n\in \{1,2,3\}\), undergoing a motion characterized by a deformation mapping $\vec{\varphi}:\mathcal{B}_{0} \rightarrow \mathbb{R}^n$,  $\vec{X}\mapsto\vec{x} = \vec{\varphi}(\vec{X})$, which maps material points $\vec{X} \in \mathcal{B}_{0}$ of the reference configuration $\mathcal{B}_{0}$ onto points $\vec{x} \in \mathcal{B}_{t}$ of the current configuration $\mathcal{B}_t = \vec{\varphi}(\mathcal{B}_{0})$. The material deformation gradient is defined by $\vec{F}:\mathcal{B}_0 \rightarrow\mathbb{R}^{n\times n}$, $\vec{F} = \nablaq{\vec{X}}(\vec{x})$, which maps infinitesimal vectors $\d\vec{X}$ at the material point $\vec{X}$ to the infinitesimal vector $\d \vec{x}$ at $\vec{x}$ in the deformed configuration. An infinitesimal area element, oriented at the material point $\vec{X}$ with the outward, reference normal vector $\vec{N}$ can be defined by two linearly independent vectors via $\d\vec{X}^{(1)}\times \d\vec{X}^{(2)} = \vec{N}\d A$. The well-known Nanson's formula reads $\vec{n}\d a = \op{cof}(\vec{F})\vec{N}\d A$,  where $\op{cof}(\vec{F}) = \vec{H}$ is the co-factor,  given by 
\begin{equation}
\vec{H} := \frac{1}{2}(\vec{F}\dcross\vec{F}).
\end{equation}
Here, we apply the cross product of two second-order tensors $\vec{A}$ and $\vec{B}$ with $[\vec{A}\dcross\vec{B}]_{iJ} = \epsilon_{ijk}\epsilon_{JMN}[\vec{A}]_{jM}[\vec{B}]_{kN}$, using the third-order Levi-Civita permutation symbol $\epsilon_{ijk}$. The infinitesimal volume element $\d V$ in the material configuration is related to the deformed counterpart via the Jacobian $J \!:=\!\op{det}[\vec{F}]>0$, where the determinant is defined by
\begin{equation}
\op{det}(\vec{F}) := \frac{1}{6}\vec{F}:(\vec{F}\dcross\vec{F}).
\end{equation}
A graphical representation of the different kinematical values is given in Fig.\ \ref{fig:kin}, see \cite{BETSCH2018660} for further details.

Introducing the space of virtual or admissible test functions for the deformation 
\begin{equation}
\mathcal{V}^{\delta\vec{\varphi}} = \{\delta\vec{\varphi}\in {H}^1(\mathcal{B}_0)\,|\,\delta\vec{\varphi}=\vec{0}\,\text{on}\,\Gamma^{\varphi}\},
\end{equation}
where $\Gamma^{\varphi}\in\partial\mathcal{B}_0$ refers to the Dirichlet boundary,  the internal virtual work  reads
\begin{equation}\label{eq:internal}
\delta\Pi^{\mathrm{int},\varphi} = \int\limits_{\mathcal{B}_0}\vec{P}:\nablaq{\vec{X}}(\delta\vec{\varphi})\,\d V,
\end{equation}
where $\vec{P}$ denotes the first Piola-Kirchhoff stress tensor.  Applying integration by parts and equating the resulting terms with external contributions yields 
\begin{equation}\label{eq:extern}
\begin{aligned}
\nablaq{\vec{X}}\cdot\vec{P}^\mathrm{T} + \vec{B} &= 0,\\
\vec{PN} - \vec{T} &= 0,\\
\vec{\varphi} - \bar{\vec{\varphi}} &= 0,
\end{aligned}
\end{equation}
where $\vec{B}$ is a body- and $\vec{T}$ the Neumann surface load on $\Gamma^\sigma\in \partial\mathcal{B}_0$, noting that 
$\Gamma^\varphi\cap\Gamma^\sigma = \emptyset$ and $\bar{\Gamma}^\varphi\cup\bar{\Gamma}^\sigma = \partial\mathcal{B}_0$. 

Without loss of generality of the chosen approach, we restrict ourselves here to elastic materials, assuming that ductile and general inelastic models can be applied as well. To describe elastic materials in the large strain regime,  the concept of polyconvexity as a mathematically well-accepted requirement must be satisfied by the strain energy density function, see \cite{bonet2015a} for details. In particular, we require the strain energy density function $\Psi$ to be a convex multi-valued and inhomogeneous function given by
\begin{equation}\label{eq:constitutive}
\Psi := \Psi(\vec{F}) = \bar{\Psi}(\vec{F},\vec{H},J,\vec{X}),
\end{equation}
where $\bar{\Psi}$ is convex with respect to the $2n^2+1$ values of $\vec{F}$, $\vec{H}$ and $J$. For isotropic materials such as 
a compressible Moonley-Rivlin material we can assume the constitutive relation to take the form
\begin{equation}\label{eq:MR}
\bar{\Psi}_{MR} := \alpha(\vec{X})\vec{F}:\vec{F} + \beta(\vec{X})\vec{H}:\vec{H} + f(J,\vec{X}), 
\end{equation}
where $\alpha$ and $\beta$ are positive material parameters locally defined at $\vec{X}$ and $f$ is a convex function of $J$ and $\vec{X}$.  The first Piola-Kirchhoff stress tensor follows immediately from
\begin{equation}
\vec{P} := \frac{\partial \Psi}{\partial\vec{F}} = \frac{\partial \bar{\Psi}}{\partial\vec{F}} +  \frac{\partial\bar{\Psi}}{\partial\vec{H}}\dcross\vec{F} + \frac{\partial\bar{\Psi}}{\partial J}\vec{H}.
\end{equation}
With regard to \eqref{eq:extern},  we obtain for the virtual work of the internal and external contributions
\begin{equation}\label{eq:weakMech}
G^{\vec{\varphi}} := \int\limits_{\mathcal{B}_0}\vec{P}:\nablaq{\vec{X}}(\delta\vec{\varphi})\,\d V - \int\limits_{\mathcal{B}_0}\delta\vec{\varphi}\cdot\vec{B}\, \d V -
\int\limits_{\Gamma^\sigma} \delta\vec{\varphi}\cdot\vec{T}\, \d A,
\end{equation} 
and require $G^{\vec{\varphi}} = 0$ for all $\delta\varphi\in\mathcal{V}^{\delta\vec{\varphi}}$. 

\subsection{Micro-continuum}
To account for the influence of the underlying microstructure on the material behaviour, we introduce a representative elementary volume (RVE) 
$\RVE\subseteq\mathbb{R}^n$, link it to each point $\vec{X}\in\mathcal{B}_0$ via a first-order Taylor approximation%
\footnote{As shown in \cite{hesch2022}, we need a higher-order Taylor approximation in the case of higher-order problems.} 
\begin{equation}\label{eq:microdefo}
\tilde{\vec{\varphi}}(\vec{X},\tilde{\vec{X}}) = \vec{F}(\vec{X})\tilde{\vec{X}} + \tilde{\vec{w}}(\vec{X},\tilde{\vec{X}}),
\quad\vec{X}\in\mathcal{B}_0,\,\tilde{\vec{X}}\in\RVE,
\end{equation} 
and interpret $\tilde{\vec{\varphi}}(\vec{X},\cdot):\RVE\rightarrow \mathbb{R}^n$ as  
a deformation map on $\RVE$. The associated quantities like the corresponding deformation gradient are derived analogous to the previous section. 
To distinguish between macro- and micro-entities, all quantities of the micro-continuum will be marked with a 
superimposed tilde. 

Per definition, the micro deformation consists of a homogeneous part $\vec{F}(\vec{X})\tilde{\vec{X}}$ and a non-homogeneous 
field $\tilde{\vec{w}}\colon\mathcal{B}_0\times\RVE\to\mathbb{R}^n$ referred to as microscopic fluctuations. According to Eq.\ \eqref{eq:microdefo}, 
we obtain 
\begin{equation}\label{eq:microGrad}
\tilde{\vec{F}} = \nablaq{\tilde{\vec{X}}}(\tilde{\vec{\varphi}}) = 
\vec{F}(\vec{X}) + \tilde{\vec{F}'},\quad \tilde{\vec{F}'}:=\nablaq{\tilde{\vec{X}}}(\tilde{\vec{w}}),
\end{equation}
so that for homogeneous materials with zero fluctuations, $\vec{F}(\vec{X})$ is recovered for the micro deformation gradient.

The material of $\RVE$ is described by a polyconvex strain energy density function $\tilde{\Psi}$ and we assume that the macroscopic and microscopic gradients, $\vec{F}$ and $\tilde{\vec{F}}$, as well as the elastic potentials, $\Psi$ and $\tilde{\Psi}$, are related via the volume averages
\begin{equation}\label{eq:homogenization}
\vec{F}=\frac{1}{|\RVE|}\int\limits_{\RVE}\tilde{\vec{F}}\,\d \tilde{V}\quad\text{and}\quad
\Psi=\frac{1}{|\RVE|}\int\limits_{\RVE}\tilde{\Psi}\,\d \tilde{V},
\end{equation}
where $|\RVE|$ is the volume of $\RVE$. In addition, insertion of \eqref{eq:microGrad} into \eqref{eq:homogenization}$_1$ yields the constraint
\begin{equation}\label{eq:fluctuCon}
\vec{F}=\vec{F}+\frac{1}{|\RVE|}\int_{\RVE} \nablaq{\tilde{\vec{X}}}(\tilde{\vec{w}})\,\d \tilde{V}
\;\Longleftrightarrow\;
\int\limits_{\RVE}\nablaq{\tilde{\vec{X}}}(\tilde{\vec{w}})\,\d \tilde{V} = \vec{0}
\end{equation} 
for the superimposed deformation field $\tilde{\vec{w}}$.  Applying Gauss divergence theorem, the constraint \eqref{eq:fluctuCon} can be alternatively 
expressed as 
\begin{equation}\label{eq:boundaryA}
\int\limits_{\partial\RVE}\tilde{\vec{w}}\otimes\tilde{\vec{N}}\,\d \tilde{A} = \vec{0},
\end{equation}
showing that \eqref{eq:fluctuCon} is satisfied by the alternative conditions
\begin{equation}\label{eq:Con}
\text(i)\;   \tilde{\vec{w}}=\vec{0}\quad\text{in }\RVE, \qquad
\text(ii)\;  \tilde{\vec{w}}=\vec{0}\quad\text{on }\partial\RVE, \qquad 
\text(iii)\; \tilde{\vec{w}}^+=\vec{w}^-\quad\text{on }\partial\RVE,
\end{equation} 
cf.\ Miehe~\cite{miehe_computational_1999-1}. The trivial condition $\text{(i)}$ enforces a homogeneous deformation of the entire domain and represents the basic assumption of a Voight-type homogenization (see Taylor \cite{taylor1938}). Condition $\text{(ii)}$ demands homogeneous deformations on the boundary and $\text{(iii)}$ addresses periodic boundary conditions. Assuming a static equilibrium state of the micro-continuum governed by the field equation
\begin{equation}\label{eq:microStat}
\nablaq{\tilde{\vec{X}}}\cdot \tilde{\vec{P}^\mathrm{T}} = \vec{0},\quad \text{in }\RVE,
\end{equation} 
with Eq.\ \eqref{eq:homogenization}$_2$ we obtain a relationship between macro- and micro-stress as follows  
\begin{equation}\label{eq:averageStress}
\vec{P}=\frac{\partial\Psi}{\partial\vec{F}} = 
\frac{1}{|\RVE|}\int\limits_{\RVE}\frac{\partial\tilde{\Psi}}{\partial \vec{F}}\,\d \tilde{V} =
\frac{1}{|\RVE|}\int\limits_{\RVE}\frac{\partial\tilde{\Psi}}{\partial \tilde{\vec{F}}}
\frac{\partial\tilde{\vec{F}}}{\partial \vec{F}}\,\d \tilde{V} = 
\frac{1}{|\RVE|}\int\limits_{\RVE}\tilde{\vec{P}}\,\d \tilde{V},
\end{equation}
where we have utilized in the last equation that according to \eqref{eq:microGrad},  $\partial\tilde{\vec{F}}/\partial\vec{F}$ is the identity. 

\begin{remark}
Each of the three conditions \eqref{eq:Con}, together with \eqref{eq:microStat}, ensures that the macro-homogeneity condition
\begin{equation}\label{eq:hillA}
\frac{1}{|\RVE|}\int\limits_{\RVE}\tilde{\vec{P}}:\delta\tilde{\vec{F}}\,\d \tilde{V} = 
\vec{P}:\delta\vec{F},
\end{equation} 
also known as Hill-Mandel criterion, is fulfilled. The virtual work applied in a specific point $\vec{X}\in\mathcal{B}_0$ to the system thus corresponds to the volumetric average of the virtual work in $\RVE$. In a different interpretation,  
the Hill-Mandel criterion along with the kinematic constraint in \eqref{eq:fluctuCon} justifies the use of the volume averages of the strain energy function \eqref{eq:homogenization}$_2$ and subsequently the volume averages of the micro stresses in \eqref{eq:averageStress}.
\end{remark}

\subsection{Variational formulation}
Taking into account relations \eqref{eq:homogenization}$_2$ and \eqref{eq:microGrad} between the elastic potentials, $\Psi$ and $\tilde{\Psi}$, and the gradients, $\vec{F}$, $\tilde{\vec{F}}$ and $\tilde{\vec{F}'}$, the macro-deformation and micro-fluctuation are determined as minimizer of the system's total energy, i.e.\ $\vec{\varphi}$ and $\tilde{\vec{w}}$ have to fulfil the condition
\begin{equation}\label{eq:min1}
(\vec{\varphi},\tilde{\vec{w}})=\arginf_{(\vec{u},\vec{v})\in\mathcal{V}^{\vec{\varphi}}\times\mathcal{V}^{\tilde{\vec{w}}}}\left\{\Pi^\mathrm{int}(\vec{u},\vec{v})-\Pi^\mathrm{ext}(\vec{u})\right\},
\end{equation}
where $\mathcal{V}^{\vec{\varphi}}$ and $\mathcal{V}^{\tilde{\vec{w}}}$ are suitable spaces of admissible solutions. The internal and external energy contributions are given by
\begin{equation}
\Pi^\mathrm{int}(\vec{u},\vec{v}) = \int\limits_{\mathcal{B}_0}\frac{1}{|\RVE|}\int\limits_{\RVE} 
\tilde{\Psi}[\tilde{\vec{F}}(\vec{F}[\vec{u}],\tilde{\vec{F}'}[\vec{v}])]\,\d\tilde{V}\,\d V,\quad
\Pi^\mathrm{ext}(\vec{u}) = \int\limits_{\mathcal{B}_0}\vec{B}\cdot\vec{u}\,\d V+
\int\limits_{\Gamma^{\sigma}} \vec{T}\cdot\vec{u}\, \d A.
\end{equation}
It should be noted that according to the assumptions in the previous section, no external contributions from the micro-continuum need to be considered. Furthermore, it is assumed here that both the body load and the Neumann surface load are independent of the deformation for the ease of presentation. Introducing the spaces of virtual deformations and fluctuations
\begin{equation}
\begin{aligned}
\mathcal{V}^{\delta\vec{\varphi}} &= \{\delta\vec{\varphi}\in H^1(\mathcal{B}_0)\,|\,\delta\vec{\varphi} = \vec{0} \;\text{on}\; \Gamma^{\varphi}\}, \\
\mathcal{V}^{\delta\tilde{\vec{w}}} &= \{\delta\tilde{\vec{w}}\in L_2(\mathcal{B}_0;H^1(\RVE))\,|\,\delta\tilde{\vec{w}} = \vec{0} \;\text{on}\; \delta\Omega_0\}, 
\end{aligned}
\end{equation}
standard variational calculus finally yields the variational problem: 

Find $(\vec{\varphi},\tilde{\vec{w}})\in \mathcal{V}^{\vec{\varphi}}\times\mathcal{V}^{\tilde{\vec{w}}}$, so that for all $(\delta\vec{\varphi},\delta\tilde{\vec{w}})\in\mathcal{V}^{\delta\vec{\varphi}}\times\mathcal{V}^{\delta\tilde{\vec{w}}}$
\begin{equation}\label{eq:varProb}
\begin{aligned}
\int\limits_{\mathcal{B}_0}\frac{1}{|\RVE|}\int\limits_{\RVE} \tilde{\vec{P}}:\nablaq{\vec{X}}(\delta\vec{\varphi})\,\d\tilde{V}\,\d V +
\int\limits_{\mathcal{B}_0} \frac{1}{|\RVE|}\int\limits_{\RVE} \tilde{\vec{P}}:\nablaq{\tilde{\vec{X}}}(\delta\tilde{\vec{w}})\,\d \tilde{V}\,\d V =& \\
\int\limits_{\mathcal{B}_0}\vec{B}\cdot\delta\vec{\varphi}\,\d V + \int\limits_{\Gamma^{\sigma}} \vec{T}\cdot \delta\vec{\varphi}\, \d A,\quad \forall\,&
(\delta\vec{\varphi},\delta\tilde{\vec{w}})\in\mathcal{V}^{\delta\vec{\varphi}}\times\mathcal{V}^{\delta\tilde{\vec{w}}},
\end{aligned}
\end{equation}
or equivalently
\begin{align}
\int\limits_{\mathcal{B}_0}\frac{1}{|\RVE|}\int\limits_{\RVE} \tilde{\vec{P}}\,\d\tilde{V}:\nablaq{\vec{X}}(\delta\vec{\varphi})\,\d V =
\int\limits_{\mathcal{B}_0}\vec{B}\cdot\delta\vec{\varphi}\,\d V + \int\limits_{\Gamma^{\sigma}} \vec{T}\cdot \delta\vec{\varphi}\, \d A,\quad \forall\,\delta\vec{\varphi}\in\mathcal{V}^{\delta\vec{\varphi}},\\
\int\limits_{\mathcal{B}_0} \frac{1}{|\RVE|}\int\limits_{\RVE} \tilde{\vec{P}}:\nablaq{\tilde{\vec{X}}}(\delta\tilde{\vec{w}})\,\d \tilde{V}\,\d V = 0 \quad \forall\,\delta\tilde{\vec{w}}\in \mathcal{V}^{\delta\tilde{\vec{w}}}.\label{eq:local}
\end{align}
Note, that the last equation \eqref{eq:local} automatically fulfils the required field equation in \eqref{eq:microStat} with regard to the boundary conditions \eqref{eq:Con}, i.e.\ we obtain a static equilibrium of the micro-continuum.

\section{Discrete multiscale boundary value problem}\label{sec:multi}
To achieve a numerical solution for the problem, we apply a finite element framework to solve for $\vec{\varphi}$ and $\tilde{\vec{w}}$. In particular, we consider a standard displacement-based finite element approach, where we first introduce finite dimensional approximations of $\vec{\varphi}$ and $\delta\vec{\varphi}$, so that
\begin{equation}
\vec{\varphi}^{\indi{h}}(\vec{X}) = \sum\limits_{A\in\omega}N^A(\vec{X})\vec{q}_A\quad\text{and}
\quad\delta\vec{\varphi}^{\indi{h}}(\vec{X}) = \sum\limits_{A\in\omega}N^A(\vec{X})\delta\vec{q}_A,\quad \text{with}\quad \vec{X}\in\mathcal{B}_0^{\indi{h}},
\end{equation}
where $\mathcal{B}_0^{\indi{h}}$  represents a (possibly approximate) parameterization of $\mathcal{B}_0$ according to the isoparametric concept. 
Here, $\omega=\{1,\,\hdots,\,n_{\text{node}}\}$, such that $\vec{q}_A\in\mathbb{R}^n$ denotes the position vector of node $A$ and 
$N^A:\mathcal{B}_0^{\indi{h}}\to\mathbb{R}$ are global shape functions. Inserting the discrete approach in \eqref{eq:weakMech} gives  
\begin{equation}\label{eq:linSys}
\delta\vec{q}_A\cdot\int\limits_{\mathcal{B}_0^{\indi{h}}}\vec{P}^{\indi{h}}(\vec{X},\tilde{\vec{X}})\nablaq{\vec{X}}(N^A(\vec{X}))\d V  =  \delta\vec{q}_A\cdot\vec{F}^{\op{ext},A},
\end{equation}
which has to hold for every $\delta\vec{q}_A$, $A=1,\ldots,n_{\text{node}}$. Here, $\vec{F}^{\op{ext},A}$ is the nodal force vector of the external contributions, given by 
\begin{equation}
\vec{F}^{\op{ext},A} = \int\limits_{\mathcal{B}_0^{\indi{h}}}N^A\vec{B}\d V + \int\limits_{\Gamma^{\sigma,\indi{h}}}N^A\,\vec{T}\d A.
\end{equation} 
Applying a suitable quadrature formula yields for every $\delta\vec{q}_A$
\begin{equation}\label{eq:linSys2}
\sum_{k=1}^{n_\mathrm{ip}} \eta_k\left[\vec{P}^{\indi{h}}(\vec{X}_\mathrm{ip}^{k},\tilde{\vec{X}})\nablaq{\vec{X}}N^A(\vec{X}_\mathrm{ip}^{k})\right] = \vec{F}^{\op{ext},A}_{\text{quad}},
\end{equation}
where $\vec{F}^{\op{ext},A}_{\text{quad}}$ represents the quadrature of the external contributions and $\eta_k$ the integration weights. 

To obtain $\vec{P}^{\indi{h}}(\vec{X}_\mathrm{ip}^{k},\tilde{\vec{X}})$, a suitable discretization of the microscopic fluctuations $\tilde{\vec{w}}$, 
defined on $\mathcal{B}_0^{\indi{h}}\times\RVE$, is required. Therefore, we introduce the tensor product of macro- and microscale 
shape functions as follows
\begin{equation}
\begin{aligned}
\tilde{\vec{w}}^{\indi{h}}(\vec{X},\tilde{\vec{X}}) &= \sum\limits_{b\in\tilde{\omega}_1}
\sum\limits_{c\in\tilde{\omega}_2}R^b(\vec{X})\tilde{R}^c(\tilde{\vec{X}})
\tilde{\vec{w}}_{b,c} = 
\sum\limits_{B\in\tilde{\omega}}\bar{R}^B(\vec{X},\tilde{\vec{X}})\tilde{\vec{w}}_B\quad\text{and}\\
\delta\tilde{\vec{w}}^{\indi{h}}(\vec{X},\tilde{\vec{X}}) &= 
\sum\limits_{b\in\tilde{\omega}_1}\sum\limits_{c\in\tilde{\omega}_2}
R^b(\vec{X})\tilde{R}^c(\tilde{\vec{X}})\delta\tilde{\vec{w}}_{b,c} = 
\sum\limits_{B\in\tilde{\omega}}\bar{R}^B(\vec{X},\tilde{\vec{X}})\delta\tilde{\vec{w}}_B, \quad
\vec{X}\in\mathcal{B}_0^{\indi{h}},\;\tilde{\vec{X}}\in\RVE^{\indi{h}},
\end{aligned}
\end{equation}
where the sets $\tilde{\omega}_i=\{1,\ldots,n_{\mathrm{micro}}^i\}$, $i=1,2$, summarize the tensor indices describing 
the position in the tensor product structure and the natural scheme $B(b,c)=(c-1)n_{\mathrm{micro}}^1+b$ is used for 
the global numbering $\tilde{\omega}=\{1,\ldots,n_{\mathrm{micro}}\}$, with  $n_{\mathrm{micro}}=n_{\mathrm{micro}}^1
n_{\mathrm{micro}}^2$.

As usual in multiscale techniques, we apply for the macroscale basis $\{R^b\}_{b=1}^{n_\mathrm{micro}^1}$ a Delta Dirac, 
\textit{that can be formally constructed as the limit of a sequence of smooth functions with compact support converging to a distribution and satisfying the so-called shifting property, i.e.}\footnote{Taken from Reali and Hughes \cite{Reali2015b}.}
\begin{equation}
\int\limits_{\mathcal{G}}f(\vec{X})\delta(\vec{X}-\vec{X}_i) \d V = f(\vec{X}_i),
\end{equation}
for every function $f$ continuous at $\vec{X}_i\in\mathcal{G}$. In particular, computational multiscale methods like FE$^2$ evaluate the microscopic fluctuations at the integration points $\vec{X}_\mathrm{ip}^b$, which is achieved by formally setting $R^b(\vec{X})=\delta(\vec{X}-\vec{X}_\mathrm{ip}^b)$, 
$b=1,\ldots n_\mathrm{ip}$ (i.e.\ $n_\mathrm{micro}^1=n_\mathrm{ip}$), such that we obtain virtual fluctuations of the form 
\begin{equation}\label{eq:fluctVaria_dirac}
\delta\tilde{\vec{w}}^{\indi{h}}(\vec{X},\tilde{\vec{X}}) = 
\sum\limits_{b=1}^{n_{\mathrm{ip}}}
\sum\limits_{c\in\tilde{\omega}_2}\delta(\vec{X}-\vec{X}_\mathrm{ip}^b)\tilde{R}^c(\tilde{\vec{X}})\delta\tilde{\vec{w}}_{b,c}.
\end{equation}
Hence, \eqref{eq:local} reads now
\begin{equation}\label{eq:local2}
\sum\limits_{b = 1}^{n_\mathrm{ip}}\,\delta\tilde{\vec{w}}_{b,c}\cdot\int\limits_{\RVE^{\indi{h}}} \tilde{\vec{P}}^{\indi{h}}(\vec{X}_\mathrm{ip}^{b},\tilde{\vec{X}})\nablaq{\tilde{\vec{X}}}
\tilde{R}^c(\tilde{\vec{X}})\,\d \tilde{V} = 0,\quad \forall\delta\tilde{\vec{w}}_{b,c}\in\mathbb{R}^n,
\end{equation}
where we multiplied the equation by the factor $|\RVE^{\indi{h}}|$. Note, that the last statement explicitly reproduces a static equilibrium with regard to the boundary conditions presented in \eqref{eq:Con} on the microscale at each macroscale integration point. 

According to the choice of the virtual fluctuations, we use Kronecker delta functions, the 
discrete analogues of the Delta-Dirac pulses, to interpolate the solutions, 
i.e.\ we introduce interpolations of the form
\begin{equation}\label{eq:fluctSolution}
\tilde{\vec{w}}^{\indi{h}}(\vec{X},\tilde{\vec{X}}) = 
\sum\limits_{b=1}^{n_{\mathrm{ip}}}
\sum\limits_{c\in\tilde{\omega}_2}\delta'(\vec{X}-\vec{X}_\mathrm{ip}^b)\tilde{R}^c(\tilde{\vec{X}})\tilde{\vec{w}}_{b,c},\quad \delta'(\vec{X}) := 
\begin{cases}
1, & \text{if } \vec{X}=\vec{0}, \\ 0, & \text{otherwise.}
\end{cases} 
\end{equation}
Insertion in \eqref{eq:linSys2} and \eqref{eq:local2} along with a suitable quadrature on the microscale yields
\begin{align}\label{eq:linSys3} 
&\sum_{b=1}^{n_\mathrm{ip}} \eta_k\left[\frac{1}{|\RVE^{\indi{h}}|}\sum\limits_{l = 1}^{n_\mathrm{ip}^{micro}}
\left[\tilde{\eta}_l\,\tilde{\vec{P}}^{\indi{h}}\left(\vec{F}^{\indi{h}}\left(\vec{X}_\mathrm{ip}^{b}\right),
\tilde{\vec{F}}^{\prime,\indi{h}}\left(\vec{X}_\mathrm{ip}^{b},\tilde{\vec{X}}_{\mathrm{ip}}^l\right)\right) \right]\nablaq{\vec{X}}N^A\left(\vec{X}_\mathrm{ip}^{b}\right)\right] = 
\vec{F}^{\op{ext},A}_{\text{quad}},\quad A\in\omega,\\
&\sum\limits_{l = 1}^{n_\mathrm{ip}^{micro}}\tilde{\eta}_l
\left[\tilde{\vec{P}}^{\indi{h}}\left(\vec{F}^{\indi{h}}\left(\vec{X}_\mathrm{ip}^{b}\right),
\tilde{\vec{F}}^{\prime,\indi{h}}\left(\vec{X}_\mathrm{ip}^{b},\tilde{\vec{X}}_{\mathrm{ip}}^l\right)\right)\nablaq{\tilde{\vec{X}}}\tilde{R}^c\left(\tilde{\vec{X}}_\mathrm{ip}^l\right)\right] 
= \vec{0}, \quad \left(b,c\right)\in \tilde{\omega}_1\times\tilde{\omega}_2,
\label{eq:linSys4}
\end{align}
where $\tilde{\eta}_l$ are the quadrature weights at the $n_{\mathrm{ip}}^{micro}$ integration point to evaluate the integral on $\RVE^{\indi{h}}$. 
\begin{remark}
Eqs.\ \eqref{eq:linSys3} and \eqref{eq:linSys4} represent the classical FE$^2$ method, evaluated in a staggered solution 
process to be discussed in the next section.
\end{remark}

Other choices to approximate $\tilde{\vec{w}}$ and $\delta\tilde{\vec{w}}$ within $\mathcal{B}_0$ using  
suitable interpolation functions are certainly possible. We refer to \cite{bonet2015a} among others, where 
discontinuous interpolation functions are introduced in the context of Hu-Washizu and Hellinger-Reissner formulations. 
As the microscopic fluctuations are applied to calculate the stresses, we obtain similar continuity requirements for both. 
The next step after the Dirac approach described above may be the use of piecewise constant approximation functions 
on the macroscale. In particular, the construction of suitable bases no longer has to be based on the quadrature 
formula used and the underlying system of equations generally has the form  
\begin{align}\label{eq:constSolution1}
&\int\limits_{\mathcal{B}_0^{\indi{h}}}\frac{1}{|\RVE^{\indi{h}}|}\int\limits_{\RVE^{\indi{h}}}
\tilde{\vec{P}}(\vec{F}^{\indi{h}}(\vec{X}),\tilde{\vec{F}}^{\prime,\indi{h}}(\vec{X},\tilde{\vec{X}}))\,\d\tilde{V}\,
\nablaq{\vec{X}}N^A(\vec{X})\,\d V = \vec{F}^{\mathrm{ext},A}, && A\in\omega,\\\label{eq:constSolution2}
&\int\limits_{\mathcal{B}_0^{\indi{h}}} \frac{1}{|\RVE^{\indi{h}}|}\int\limits_{\RVE^{\indi{h}}} 
\tilde{\vec{P}}(\vec{F}^{\indi{h}}(\vec{X}),\tilde{\vec{F}}^{\prime,\indi{h}}(\vec{X},\tilde{\vec{X}}))
\nablaq{\tilde{\vec{X}}}\bar{R}^B(\vec{X},\tilde{\vec{X}})\,\d \tilde{V}\,\d V = \vec{0}, && B\in\tilde{\omega}. 
\end{align}
Using linear or even higher-order shape functions for $\{R^b\}_{b=1}^{n_\mathrm{micro}^1}$ can be done in a straight-forward 
manner; however, we have to ensure that the shape functions are discontinuous at the element boundaries to allow for a block 
diagonal sub-matrix of the Hessian to be used within the following null-space reduction scheme.  Moreover, we have to ensure that 
we do not obtain stability issues, as a higher-order interpolation of the micro-fluctuations (and thus, of the stresses to be calculated) 
may conflict with a lower-order interpolation of the macroscale deformation map. Both equations \eqref{eq:constSolution1} and 
\eqref{eq:constSolution2} can be rewritten as $\delta\vec{q}_A\cdot\vec{R}^A_{\mathrm{macro}} = 0$ and 
$\delta\tilde{\vec{w}}_B\cdot\vec{R}^B_{\mathrm{micro}} = 0$, where $\vec{R}^A_{\mathrm{macro}}$ and $\vec{R}^B_{\mathrm{micro}}$ are the corresponding residual vector,  to be used within a Newton-Raphson iteration as shown in the next section.

\section{Solution of the multiscale problem}\label{sec:solution}
The non-linear multiscale framework at hand is solved by introducing a Newton-Raphson iteration, 
noting that further enhancement like line-search or trust-region methods can be applied as well, 
see \cite{hesch2023} for details.  In particular, we solve
\begin{equation}\label{eq:discreteSystem}
[\delta\vec{q}_A,\,\delta\tilde{\vec{w}}_B]\cdot
\begin{bmatrix}
\vec{K}^{AC} & \vec{D}^{AD} \\
\vec{E}^{BC} & \vec{L}^{BD}
\end{bmatrix}
\begin{bmatrix}
\Delta\vec{q}_C \\ \Delta\tilde{\vec{w}}_D
\end{bmatrix} = -
[\delta\vec{q}_A,\,\delta\tilde{\vec{w}}_B]\cdot
\begin{bmatrix}
\vec{R}^A_{\text{macro}} \\ \vec{R}^B_{\text{micro}}
\end{bmatrix}.
\end{equation}
Afterwards, both values are updated via $\vec{q}_A \leftarrow \vec{q}_A + \Delta\vec{q}_A$ and $\tilde{\vec{w}}_A \leftarrow \tilde{\vec{w}}_A + \Delta\tilde{\vec{w}}_A$ and the iteration restarts until $\|[\vec{R}^\mathrm{T}_{\text{macro}},\,\vec{R}^\mathrm{T}_{\text{micro}}]^\mathrm{T}\| < \epsilon_{NR}$ with a predefined stop criterion $\epsilon_{NR}$. The tangent matrix is composed of four terms, given by
\begin{equation}\label{eq:discreteSys_K}
\vec{K}^{AC} = \int\limits_{\mathcal{B}_0^{\indi{h}}}\nablaq{\vec{X}}(N^A)\cdot\frac{1}{|\RVE^{\indi{h}}|}
\int\limits_{\RVE^{\indi{h}}}\frac{\partial^2\tilde{\Psi}}{\partial\tilde{\vec{F}}\partial\tilde{\vec{F}}}\d\tilde{V}\,\nablaq{\vec{X}}(N^C)\d V
\end{equation}
and 
\begin{equation}\label{eq:discreteSys_D}
\vec{D}^{AD} = \int\limits_{\mathcal{B}_0^{\indi{h}}}\nablaq{\vec{X}}(N^A)\cdot\frac{1}{|\RVE^{\indi{h}}|}
\int\limits_{\RVE^{\indi{h}}}\frac{\partial^2\tilde{\Psi}}{\partial\tilde{\vec{F}}\partial\tilde{\vec{F}}}\nablaq{\tilde{\vec{X}}}(\bar{R}^D)\d\tilde{V}\d V.
\end{equation}
As the tangent matrix is the Hessian resulting from the condition in \eqref{eq:min1}, the matrix is 
symmetric and $\vec{E} = \vec{D}^\mathrm{T}$. Eventually, the last term is given by
\begin{equation}\label{eq:discreteSys_L}
\vec{L}^{BD} = \int\limits_{\mathcal{B}_0^{\indi{h}}}\frac{1}{|\RVE^{\indi{h}}|}
\int\limits_{\RVE^{\indi{h}}}\nablaq{\tilde{\vec{X}}}(\bar{R}^B)\cdot
\frac{\partial^2\tilde{\Psi}}{\partial\tilde{\vec{F}}\partial\tilde{\vec{F}}}\nablaq{\tilde{\vec{X}}}(\bar{R}^D)\d\tilde{V}\d V
\end{equation} 
and the corresponding  global algebraic system reads
\begin{equation}\label{eq:discreteSystem_global}
\begin{bmatrix}
\vec{K} & \vec{D} \\
\vec{E} & \vec{L}
\end{bmatrix}
\begin{bmatrix}
\Delta\vec{q} \\ \Delta\tilde{\vec{w}}
\end{bmatrix} = -
\begin{bmatrix}
\vec{R}_{\text{macro}} \\ \vec{R}_{\text{micro}}
\end{bmatrix},
\end{equation} 
where $\Delta\vec{q}$, $\Delta\tilde{\vec{w}}$ summarize the unknown node data $\Delta\vec{q}_A$, $\Delta\tilde{\vec{w}}_B$ and 
the individual blocks are composed according to Eqs.\ \eqref{eq:discreteSystem}-\eqref{eq:discreteSys_L}.

\subsection{Classical solution strategy}\label{sec:classical_sol}
In classical FE$^2$ methods as presented in, e.g., \cite{KEIP201462} for nonlinear electromechanical multiscale problems, 
a staggered scheme is applied. First, \eqref{eq:discreteSystem_global} is solved using fixed $\vec{q}$%
\footnote{Analogous to the introductory considerations, $\vec{q}\in\mathbb{R}^{n\cdot n_\mathrm{node}}$ and 
$\tilde{\vec{w}}\in\mathbb{R}^{n\cdot n_\mathrm{micro}}$ summarize in the following the node degrees 
of freedom $\vec{q}_A$ and ${\tilde{\vec{w}}_B}$ respectively.},
until $\|\vec{R}_{\text{micro}}\| < \epsilon_{NR}^{\text{micro}}$. Then, the second line of \eqref{eq:discreteSystem_global} reads
\begin{equation}\label{eq:assumption}
\vec{E}\Delta\vec{q} + \vec{L}\Delta\tilde{\vec{w}} = \vec{0}.
\end{equation}
Since the matrix $\vec{L}$ contains all tangent matrices of all RVEs, which are strictly separated, the matrix is block-wise diagonal. 
As we have to solve the linear system in the second line of \eqref{eq:discreteSystem_global} (either using direct or iterative solver), 
we can write for this step in the solution procedure
\begin{equation}\label{eq:reduction}
\Delta\tilde{\vec{w}} = -\vec{L}^{-1}\vec{E}\Delta\vec{q},
\end{equation}
and insert this in the first line of \eqref{eq:discreteSystem_global}, such that
\begin{equation}\label{eq:solution1}
\left[\vec{K} - \vec{D}\vec{L}^{-1}\vec{E}\right]\Delta\vec{q} = -\vec{R}_{\text{macro}},
\end{equation}
update $\vec{q}$ with $\Delta\vec{q}$, update $\tilde{\vec{w}}$ using \eqref{eq:reduction} and restart the staggered 
scheme again until $\|\vec{R}_{\text{macro}}\|<\epsilon_{\text{macro}}$. 

\subsection{Generalization of the solution strategy}\label{sec:generalized_sol}
The solution strategy introduced in the previous section can be rewritten as premultiplication with the rectangular 
null-space matrix
\begin{equation}
\vec{P} = \begin{bmatrix}\vec{I}, & -\vec{D}\vec{L}^{-1}\end{bmatrix}^\mathrm{T},
\end{equation}
where $\vec{I}$ is the unity matrix of dimension ${n_{\text{node}}\times n_{\text{node}}}$. 
Now, instead of applying a staggered scheme, we multiply the linear system composed of both scales using the null-space matrix
\begin{equation}
\vec{P}^\mathrm{T}\cdot
\begin{bmatrix}
\vec{K} & \vec{D} \\
\vec{E} & \vec{L}
\end{bmatrix}
\begin{bmatrix}
\Delta\vec{q} \\ \Delta\tilde{\vec{w}}
\end{bmatrix} = -\vec{P}^\mathrm{T}\cdot
\begin{bmatrix}
\vec{R}_{\text{macro}} \\ \vec{R}_{\text{micro}}
\end{bmatrix},
\end{equation}
which yields
\begin{equation}\label{eq:solution2}
\left[\vec{K}-\vec{D}\vec{L}^{-1}\vec{E}\right]\Delta\vec{q} = -\left[\vec{R}_{\text{macro}}-\vec{D}
\vec{L}^{-1}\vec{R}_{\text{micro}}\right].
\end{equation}
Finally, we need to calculate $\Delta\tilde{\vec{w}}$ via
\begin{equation}
\Delta\tilde{\vec{w}} = -\vec{L}^{-1}\left[\vec{R}_{\text{micro}}+\vec{E}\Delta\vec{q}\right],
\end{equation}
and update both values $\vec{q}$ and $\tilde{\vec{w}}$ within the Newton-Raphson iteration.

\begin{remark}
For the non-linear case,  $\|\vec{R}_{\text{micro}}\|$ is usually unequal zero due to numerical issues,  hence \eqref{eq:assumption} 
is never valid and may affect the quadratic convergence of the Newton-Raphson iteration. Thus, the proposed 
approach is in general advantageous as we can apply the staggered scheme to \eqref{eq:solution2}. In comparison 
to classical (staggered) schemes, the term $-\vec{D}\vec{L}^{-1}\vec{R}_{\text{micro}}$ on the right-hand side 
of \eqref{eq:solution2} is introduced, noting again, that $\vec{R}_{\text{micro}}$ is never exact zero. 
\end{remark}
\section{Numerical examples}\label{sec:numerics}
In this Section, we demonstrate the advantages of the proposed formulation. In particular, 
we start with a convergence study using an analytical solution of a 1+1 dimensional problem, 
followed by a typical problem emanating from non-linear elasticity.

\subsection{Convergence studies}
\begin{figure}[ht]
\begin{center}
\footnotesize
\begin{tabular}{cc}
\includegraphics[width=0.435\textwidth]{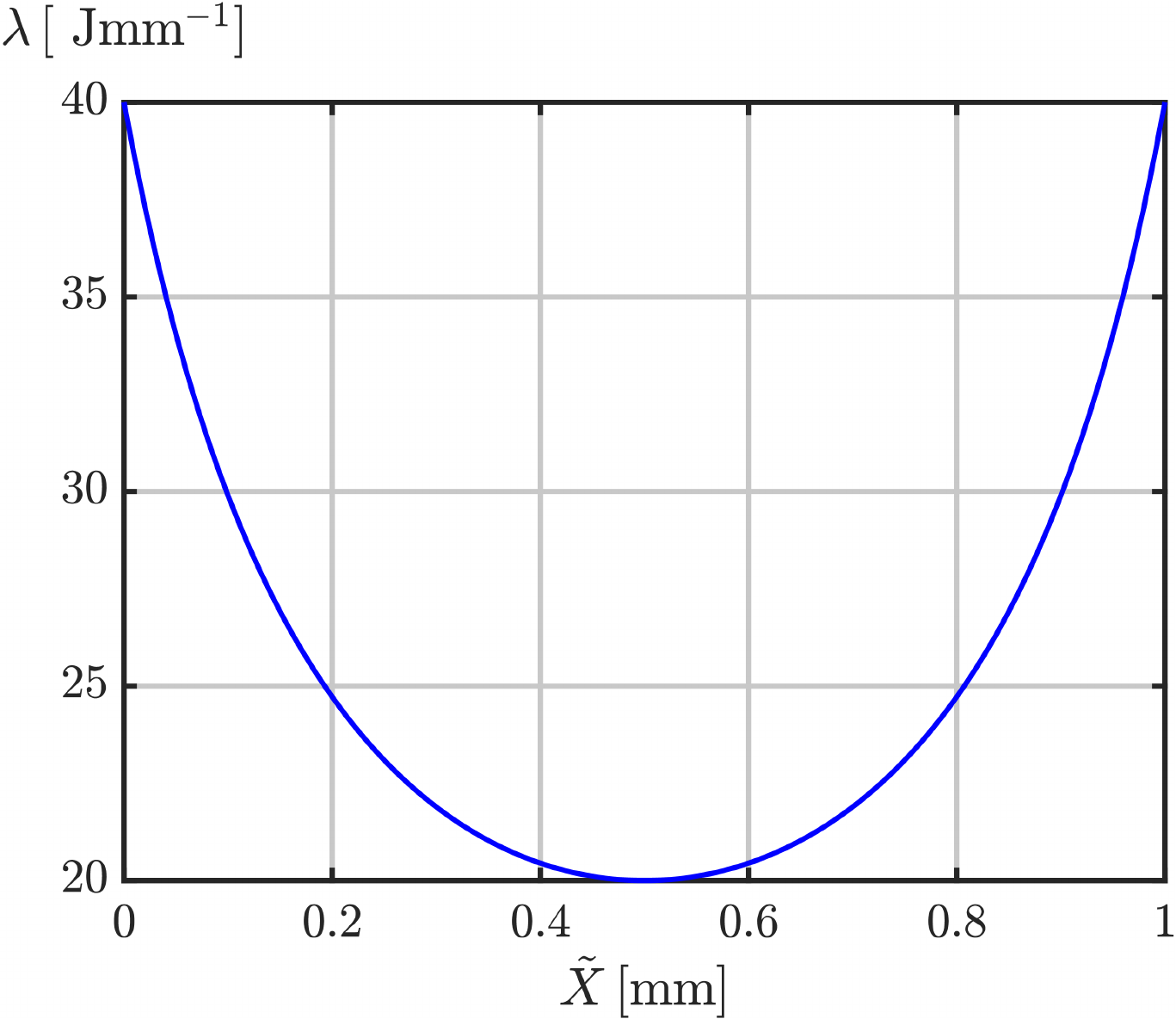} \hspace{6mm}
\includegraphics[width=0.435\textwidth]{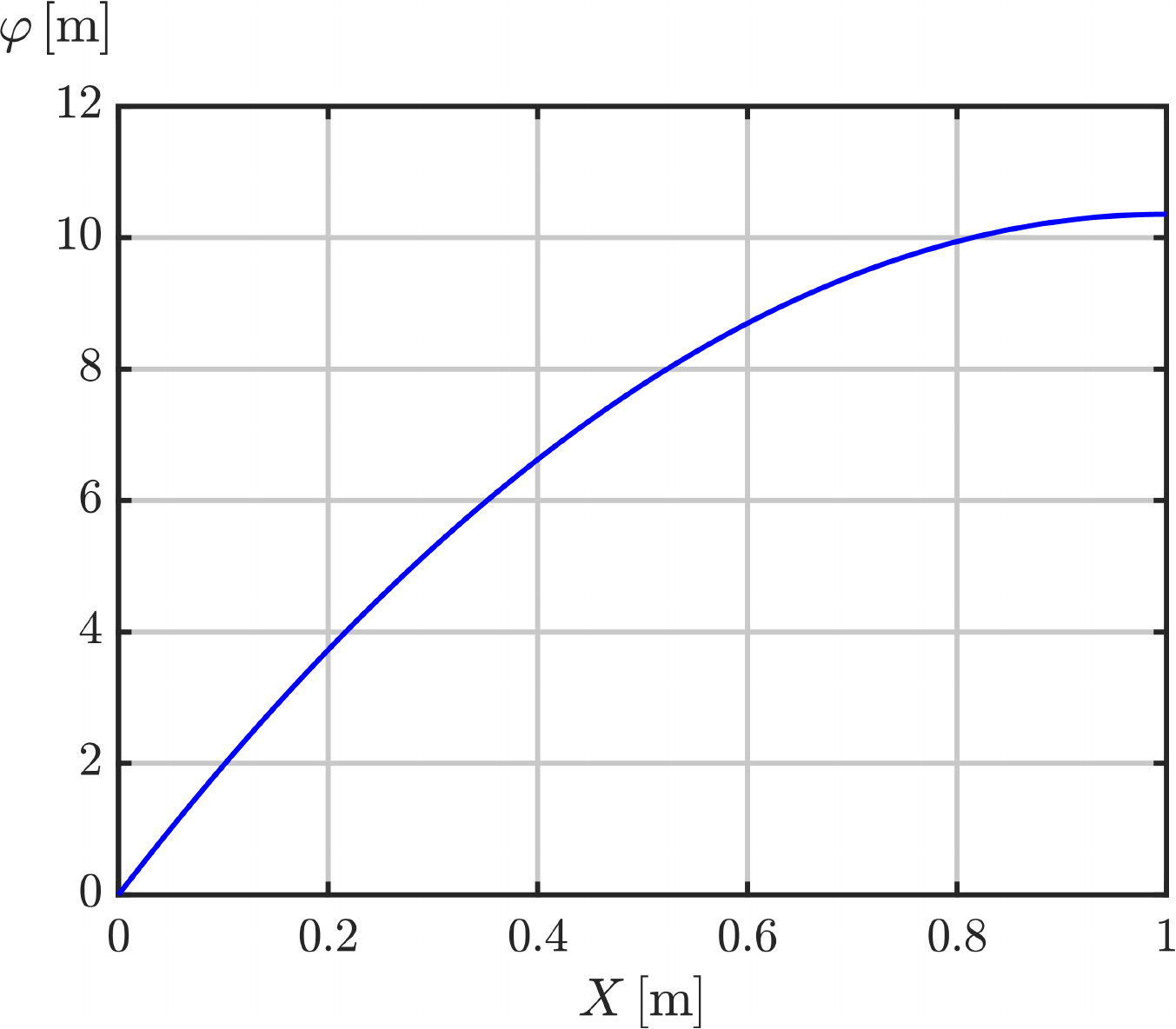} \\
\includegraphics[width=0.46\textwidth]{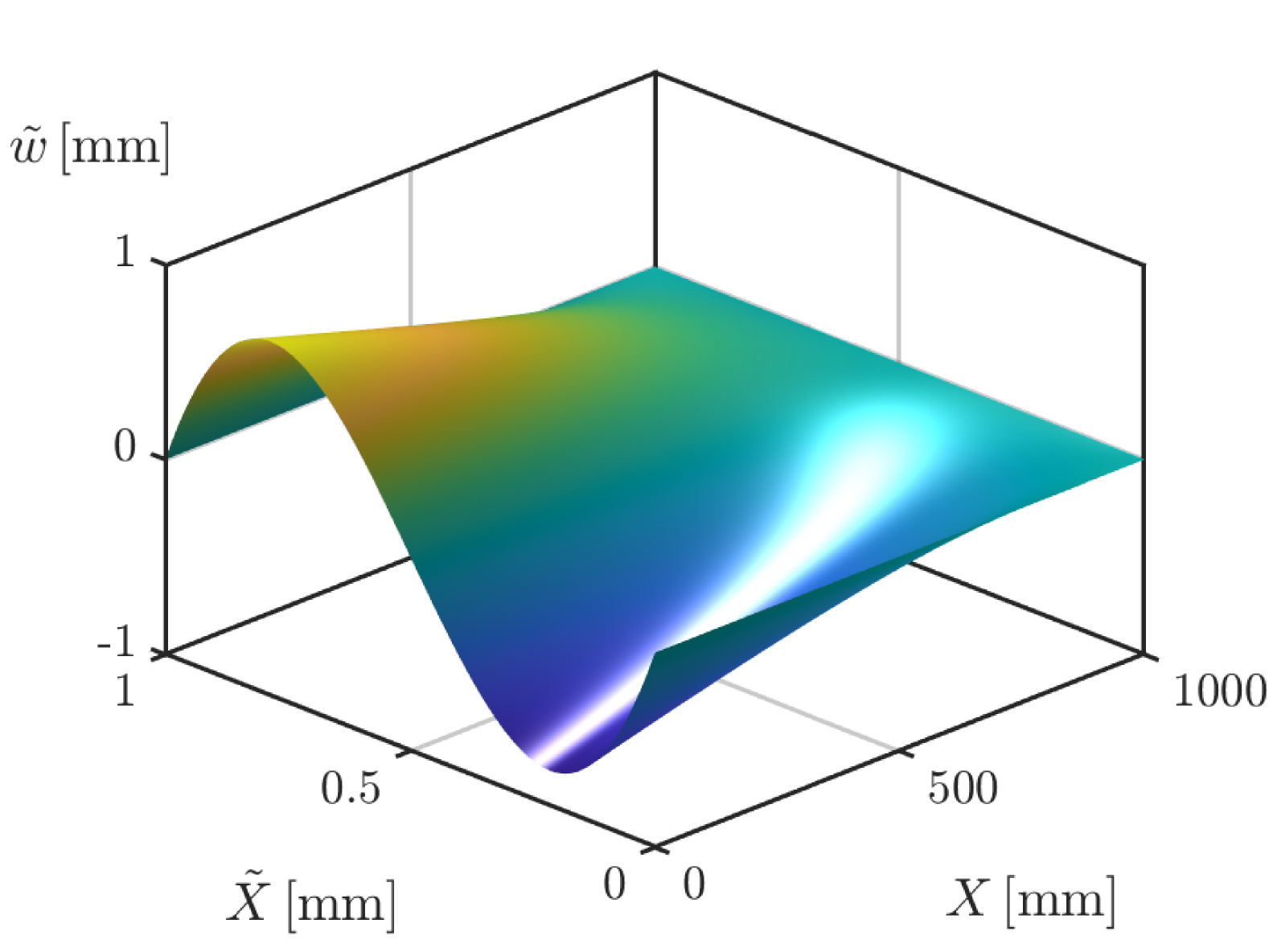}        
\includegraphics[width=0.46\textwidth]{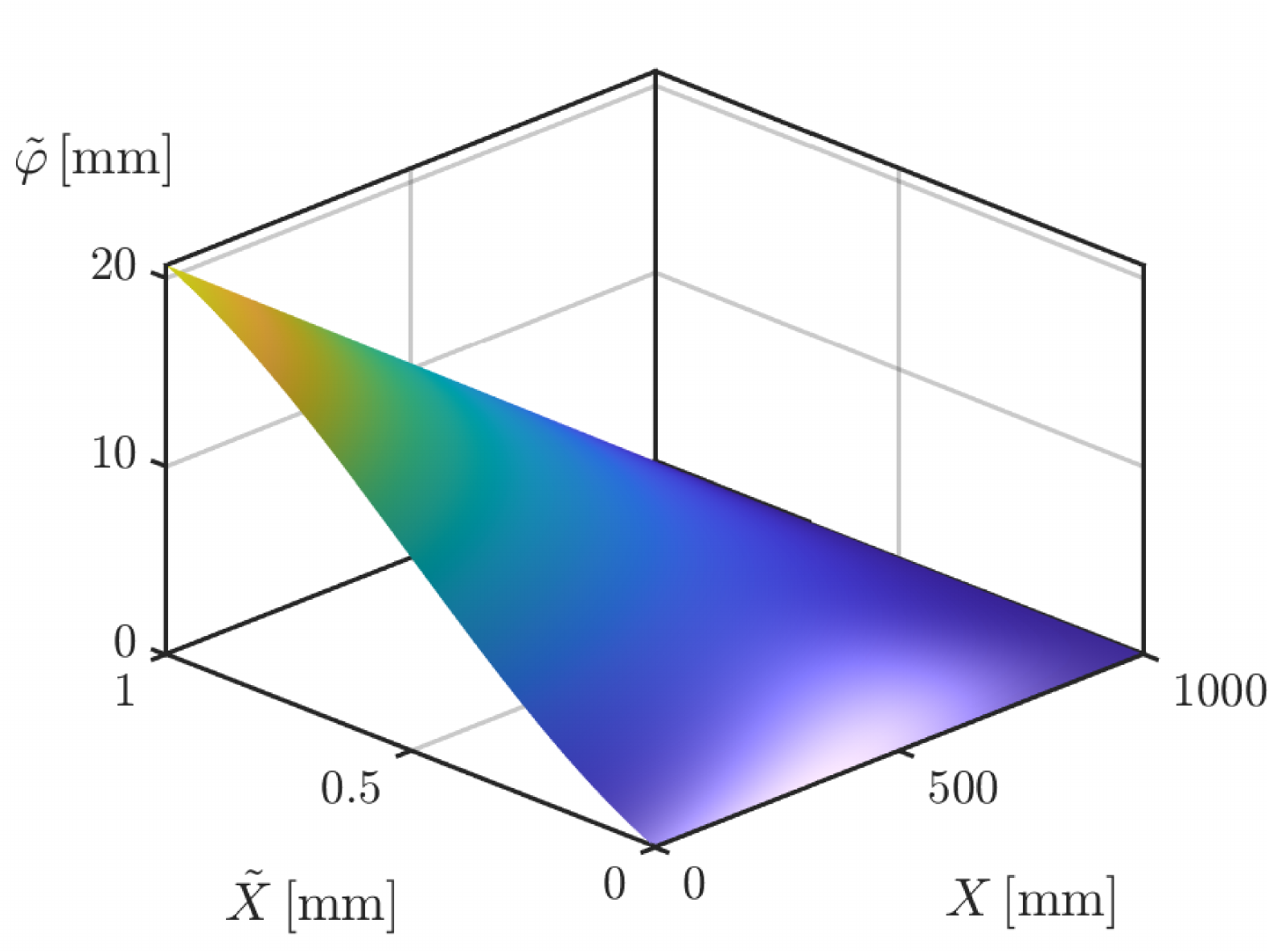} 
\end{tabular}
\end{center}
\caption{Upper left: material parameter $\lambda$, upper right: macro-deformation $\varphi$, 
lower left: micro-fluctuation $\tilde{w}$, lower right: micro-deformation $\tilde{\varphi}$.}
\label{fig:matSol}
\end{figure}

In the following, some numerical convergence analyses are presented using a one-dim\-ensional 
benchmark problem on the macroscale and a one-dimensional microscale. The macro- and micro-continuum are given by 
$\mathcal{B}_0=(0,1000)$ and $\RVE=(0,1)$, respectively, where the lengths here and in the remainder of this example are given 
in millimetres unless stated otherwise. The material is defined by the strain energy density
\begin{equation}
\tilde{\Psi}(\tilde{F},\tilde{X}) = \lambda(\tilde{X})(\tilde{F}^2-1), \quad 
\lambda(\tilde{X}) = 20\left[\cos\left(\frac{2\pi}{3}\tilde{X}-\frac{\pi}{3}
\right)\right]^{-1},
\end{equation} 
where $\lambda$ is given in the unit Jmm$^{-1}$ and represents a location-dependent material 
property, cf.\ Figure~\ref{fig:matSol}.   
It should be noted that $\tilde{\Psi}$ is selected so that an analytical solution 
of the system can be specified. Due to the simple shape, however, residual stresses occur, 
so the reference configuration is not stress-free. Assuming the body load and boundary 
conditions 
\begin{equation}
B = 1\,\text{Jmm}^{-1},\quad \varphi(0) = 0, \quad \varphi(1000) = \frac{37575\sqrt{3}}{2\pi}
\end{equation}
for the macro-deformation and homogeneous Dirichlet conditions in accordance with 
Eq.\ \eqref{eq:Con}$_{\text{ii}}$ 
for the micro-fluctuation, 
the solutions $\varphi^a\colon\mathcal{B}_0\to\mathbb{R}$ and 
$\tilde{w}^a\colon\mathcal{B}_0\times\RVE\to\mathbb{R}$ of the coupled boundary value 
problems \eqref{eq:extern}, \eqref{eq:microStat} are given by 
\begin{align}
\varphi^a(X) &= -\frac{3\sqrt{3}}{160\pi}\,X^2+\frac{3003\sqrt{3}}{80\pi}\,X,\label{eq:sol1D_phi}\\
\tilde{w}^a(X,\tilde{X}) &= F^a(X)
\left[\frac{1}{\sqrt{3}}\sin\left(\frac{2\pi}{3}\,\tilde{X}-\frac{\pi}{3}\right)-
\tilde{X}+\frac{1}{2}\right],                                                 \label{eq:sol1D_wt}
\end{align}
where $F^a=\d\varphi^a/\d X$, see Figure~\ref{fig:matSol} for illustration. 
Taking into account the given data, the two functions are characterized according to 
Eq.\ \eqref{eq:varProb} by the conditions 
\begin{align}
\int\limits_0^{1000}\int\limits_0^1 \lambda\left(\frac{\d\varphi}{\d X}+
\frac{\partial\tilde{w}}{\partial\tilde{X}}\right)\,\frac{\d \delta\varphi}{\d X}\,\d 
\tilde{X}\d X &= \frac{1}{2}\int\limits_0^{1000}\delta\varphi\,\d X, \quad 
\forall\,\delta\varphi\in\mathcal{V}^{\delta\varphi},\\
\int\limits_0^{1000}\int\limits_0^1 \lambda\left(\frac{\d\varphi}{\d X}+\frac{\partial\tilde{w}}{\partial\tilde{X}}
\right)\,\frac{\partial \delta\tilde{w}}{\partial \tilde{X}}\d \tilde{X}\d X &= 0, \quad 
\forall\,\delta\tilde{w}\in \mathcal{V}^{\delta\tilde{w}},
\end{align}
where the various carried out finite element analyses are based on. In the following, 
a total of six settings is considered, each of which differs in the choice of approximation functions. 
In detail, bilinear ($p=[1,1]$), biquadratic ($p=[2,2]$) and mixed approaches of the form $p=[0,1]$, $p=[0,2]$ are 
used for the fluctuation field, whereby in the last two cases $\tilde{w}$ is approximated by piecewise constant 
functions in $\vec{X}$-direction. In addition, approaches of the form $p=[\delta,1]$, and $p=[\delta,2]$ are considered, i.e.\ 
two classical approaches in which Dirac pulses are used in $X$-direction of $\tilde{w}$, cf.\  Eqs.\ \eqref{eq:fluctVaria_dirac} 
and \eqref{eq:fluctSolution}. Regarding the macro deformation, shape-functions of the same order as in $\tilde{X}$-direction of 
$\tilde{w}$ are used in each case.
\begin{figure}[ht]
\begin{center}
\footnotesize
\begin{tabular}{cc}
\includegraphics[width=0.435\textwidth]{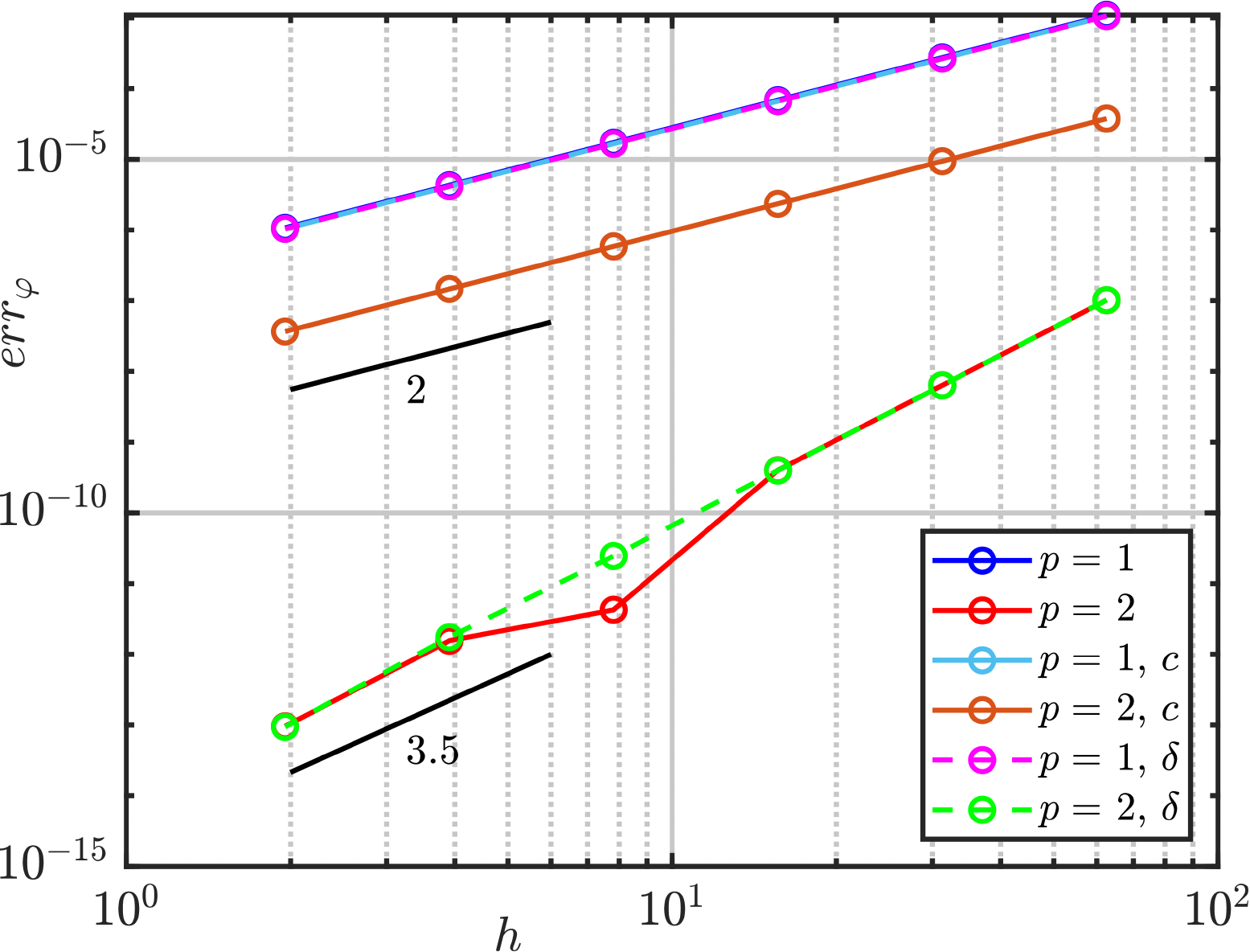} \hspace{6mm}
\includegraphics[width=0.435\textwidth]{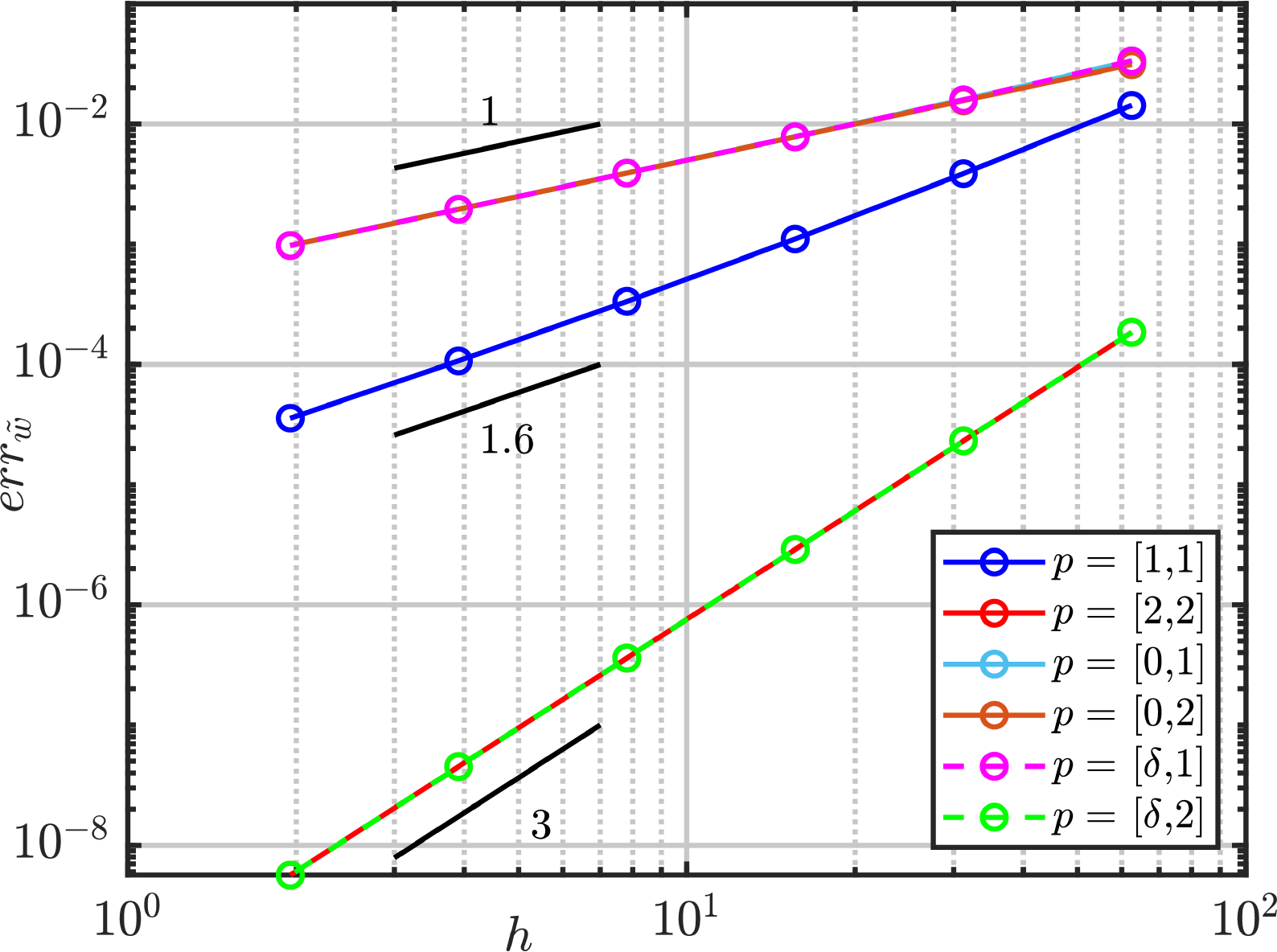}
\end{tabular}
\end{center}
\caption{Left: $err_{\varphi}$ as a function of the mesh-size $h$. The two cases in which piecewise constant shape-functions 
where used in $X$-direction of $\tilde{w}$ are labelled witch ``c'', and the cases in which Dirac pulses were used are 
labelled with ``$\delta$'' accordingly. Right: $err_{\tilde{w}}$ as a function of the mesh-size $h$. 
The error curves of $p=[0,1],\,[0,2]$ and $p=[\delta,1]$ overlap so that only one curve for these three cases is recognisable.}
\label{fig:L2err_1D}
\end{figure}
Figure~\ref{fig:L2err_1D} shows the results of a convergence study where the relative $L_2$ errors 
\begin{equation}
err_{\varphi} := \frac{\|\varphi^\mathrm{h}-\varphi^a\|_{L_2(\mathcal{B}_0)}}{\|\varphi^a\|_{L_2(\mathcal{B}_0)}},\qquad
err_{\tilde{w}} := \frac{\|\tilde{w}^\mathrm{h}-\tilde{w}^a\|_{L_2(\mathcal{B}_0\times\RVE)}}
{\|\tilde{w}^a\|_{L_2(\mathcal{B}_0\times\RVE)}}
\end{equation}
corresponding to each setting are plotted as functions of the mesh size $h$ which is defined as the maximal diameter 
of the elements in the reference configuration. Moreover, we denote by $\|\bullet\|_{L_2(B)}$ the usual $L_2$ norm on a 
domain $B$. For the calculations, resolutions of $2^k$ elements are used for the macro-domain $\mathcal{B}_0$ and 
corresponding  resolutions of $2^k\times 2^k$ elements are used for the extended region $\mathcal{B}_0\times\RVE$, 
where $k=4,\ldots,9$. As can be seen in the left-hand image of Figure~\ref{fig:L2err_1D}, $err_{\varphi}$ reduces according 
to an order greater than $2$ in the case of linear elements regardless of which approach is used for 
$\tilde{w}$ in $X$-direction. The situation is different when using quadratic approaches. While with 
an approximation of $\tilde{w}$ with $p=[2,2],\,[\delta,2]$ the error decreases with an order of  
$3.5$, an order of $2$ can be observed using the piecewise constant/quadratic approach $p=[0,2]$. 
Eventually, the behaviour of the error $err_{\tilde{w}}$ is shown in Figure~\ref{fig:L2err_1D} on the right. 
There it can be seen that a convergence order of $1.6$ is achieved with the bilinear approach, while an 
order of $3$ is achieved with both the biquadratic and the quadratic Dirac approach $p=[\delta,2]$. 
In addition, the mixed approaches $p=[0,1],\,[0,2]$ and the linear Dirac approach $p=[\delta,1]$ each 
achieve a convergence order of $1$. We point out, that the corresponding error curves overlap in the 
graphic so that only one curve can be recognized for these three cases.     
\subsection{Cook's membrane}
\begin{figure}[ht]
\begin{center}
\footnotesize
\begin{tabular}{cc}
\includegraphics[width=0.3825\textwidth]{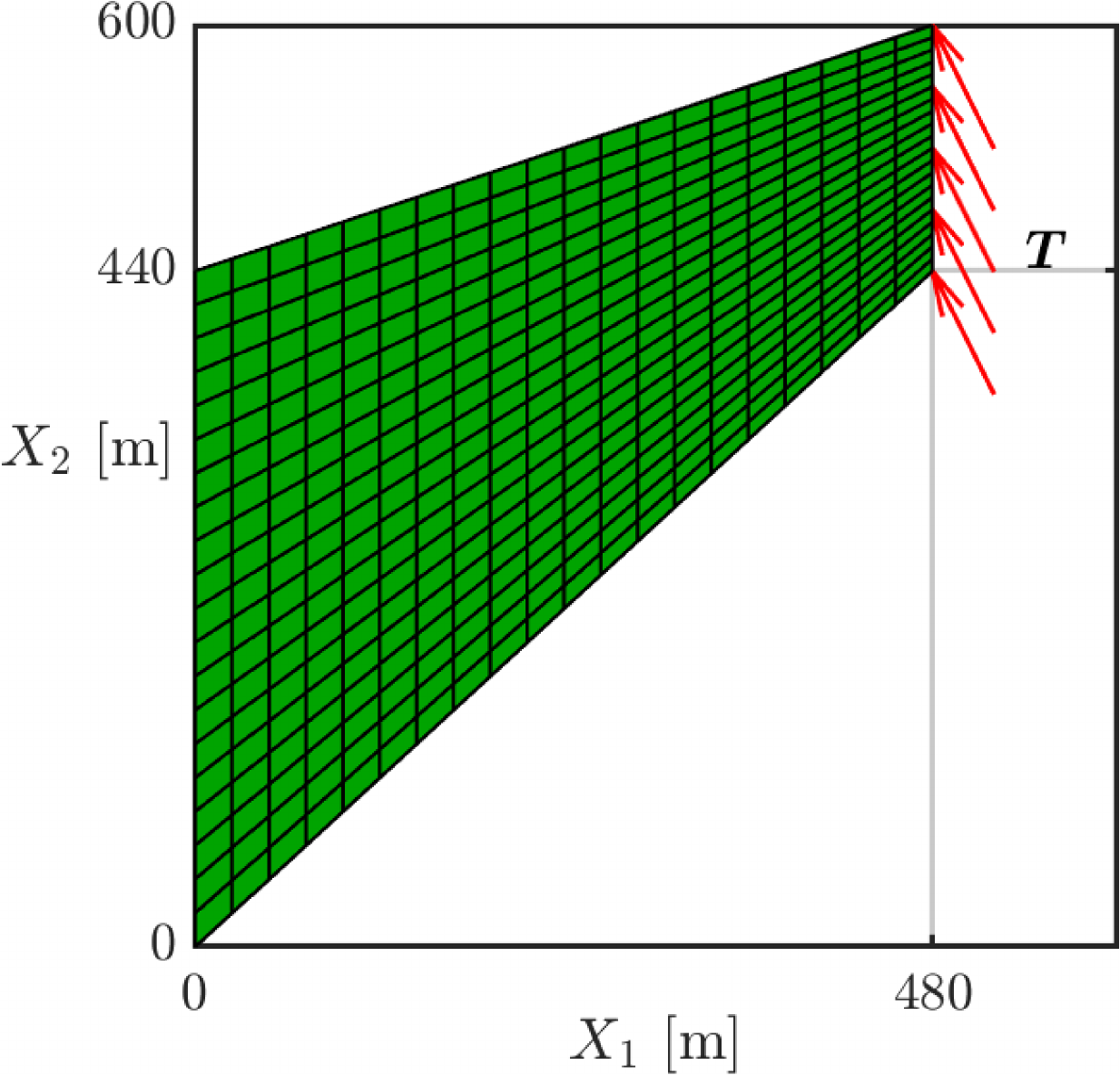}\hspace{12mm}
\includegraphics[width=0.4\textwidth]{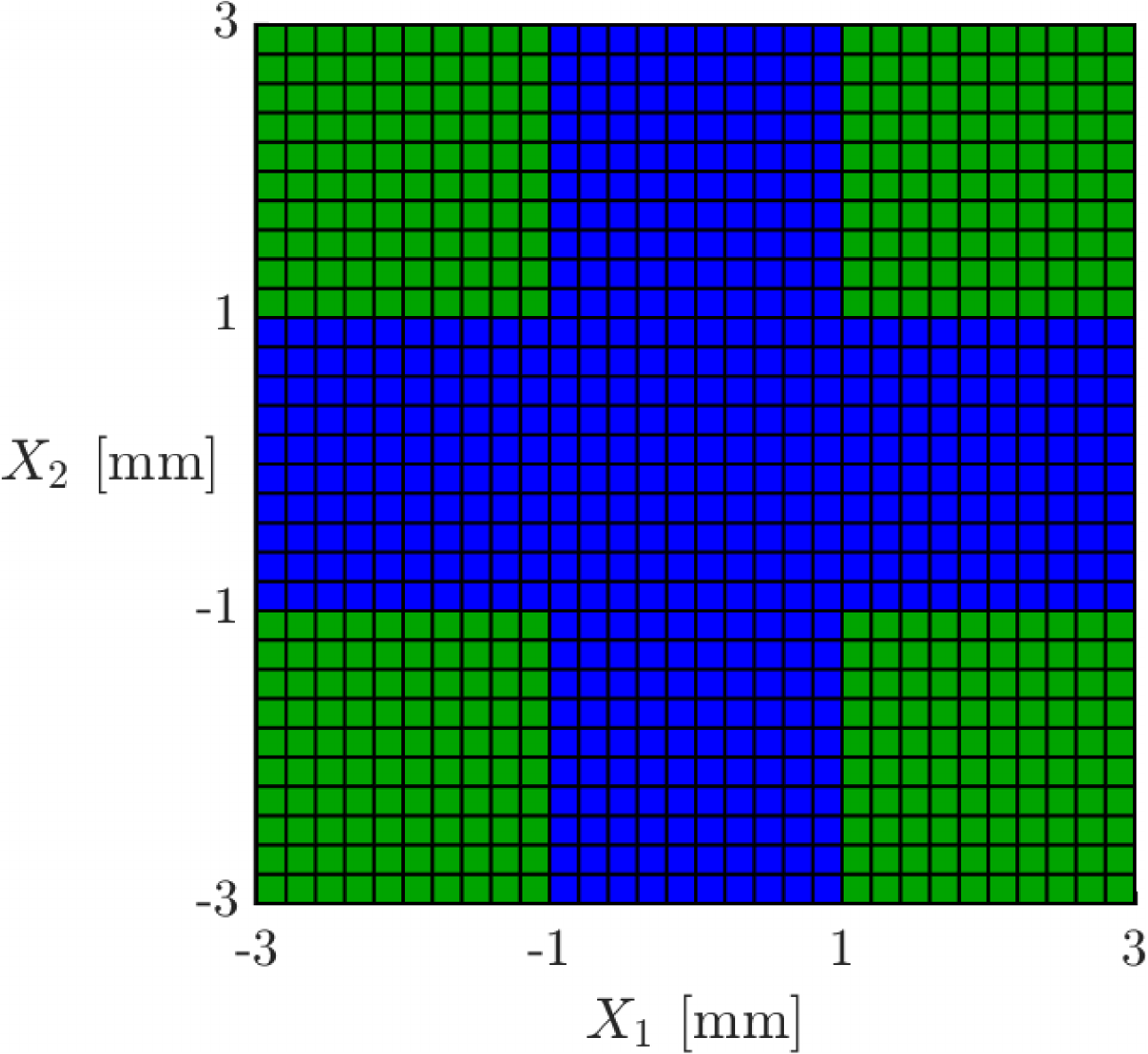}
\end{tabular}
\end{center}
\caption{Left: reference configuration and computational mesh of the macro-system. The 
red arrows indicate the impact of the surface load defined in Eq.\ \eqref{eq:load}. 
Right: reference configuration and computational mesh 
of an RVE. The green regions consist of the material defined by $\tilde{\Psi}_1$ and 
the blue one consist of the material defined by $\tilde{\Psi}_2$.}
\label{fig:refCon}
\end{figure}

Finally, we compare the two solution strategies presented in Section~\ref{sec:solution} on the basis of 
a two-dimensional Cook's membrane. The reference configuration of the macro-continuum is defined by the four 
points $\vec{P}_1=(0,0)^\mathrm{T}$, $\vec{P}_2=(480,440)^\mathrm{T}$, $\vec{P}_3=(480,600)^\mathrm{T}$, 
$\vec{P}_4=(0,440)^\mathrm{T}$, where $\vec{P}_1$ denotes the lower left, $\vec{P}_2$ the lower right, 
$\vec{P}_3$ the upper right and $\vec{P}_4$ the upper left corner of the membrane. Consequently, the computational  
macro-domain is given by
\begin{equation}\label{eq:refcon}
\mathcal{B}_0 = \left\{\vec{X}=(X_1,X_2)\in\mathbb{R}^2\, |\, 
X_1\in(0,480),\, \frac{11}{12}\,X_1<X_2<\frac{1}{3}\,X_1+440 \right\}
\end{equation}
while we use a square area of the form $\RVE=(-3,3)^2$ to represent the micro-scale.
Here and in the following, all length specifications referring to the macro-scale are given in metres 
and length specifications referring to the micro-scale are given in millimetres.
Regarding the boundary conditions, we assume that the left edge 
$\Gamma^{\varphi}=\{\vec{X}\in\mathbb{R}^2\, | \, X_1=0,\, 0<X_2<440\}$ of the membrane is fixed, 
whereas the remaining boundary $\Gamma^{\sigma}=\partial\mathcal{B}_0\backslash \Gamma^{\varphi}$ 
is exposed to the surface load 
\begin{equation}\label{eq:load}
\vec{T}\colon\ \Gamma^{\sigma}\to\mathbb{R}^2,\quad
\vec{T}(\vec{X}) = \begin{cases}
(-5,10)^\mathrm{T} & \text{if } \vec{X}\in\Gamma^{\sigma}_r, \\
(0,0)^\mathrm{T}   & \text{otherwise},
\end{cases}
\end{equation}
acting on the right edge 
$\Gamma^{\sigma}_r := \{\vec{X}\in\mathbb{R}^2\, | \, X_1=480,\, 440<X_2<600\}$ of $\mathcal{B}_0$, 
see the left picture in Figure~\ref{fig:refCon} for illustration.
Moreover, we assume a zero body load ($\vec{B}=\vec{0}$) and the membrane to consist of two materials 
described by the strain energy densities  
\begin{equation}
\tilde{\Psi}_i(\tilde{\vec{F}}) = \alpha_i(\tilde{\vec{F}}:\tilde{\vec{F}}-2)+
\beta_i(\tilde{\vec{F}}:\tilde{\vec{F}}+\tilde{J}^2-3)+\frac{\kappa_i}{2}(\tilde{J}-1)^2-
2(\alpha_i+2\beta_i)\log(\tilde{J}), 
\end{equation}
$i=1,2$, with the material parameters  $(\alpha_1,\beta_1,\kappa_1)=(27,18,60)\,$Jmm$^{-2}$ 
and $(\alpha_2,\beta_2,\kappa_2)=(13.5,6.5,30)\,$Jmm$^{-2}$, respectively. 
The two materials are distributed within the RVE in such a way that a cross-shaped area 
of width $2\,$mm consists of the second material, while the remaining area consists of the first material, 
see the right picture in Figure~\ref{fig:refCon} for illustration. 

\begin{figure}[ht]
\begin{center}
\footnotesize
\begin{tabular}{cc}
\includegraphics[width=0.3225\textwidth]{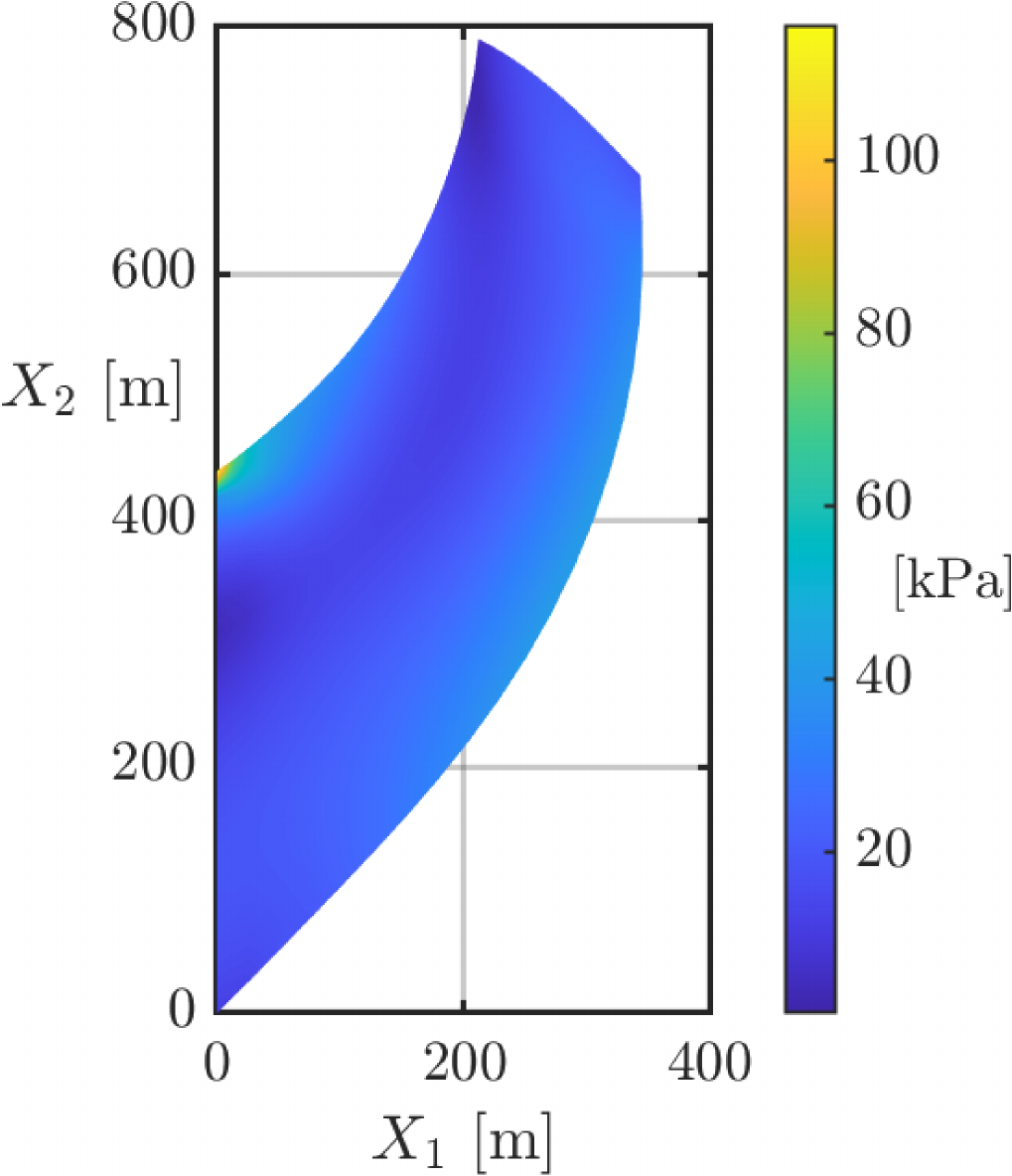}\hspace{10mm}
\includegraphics[width=0.48\textwidth]{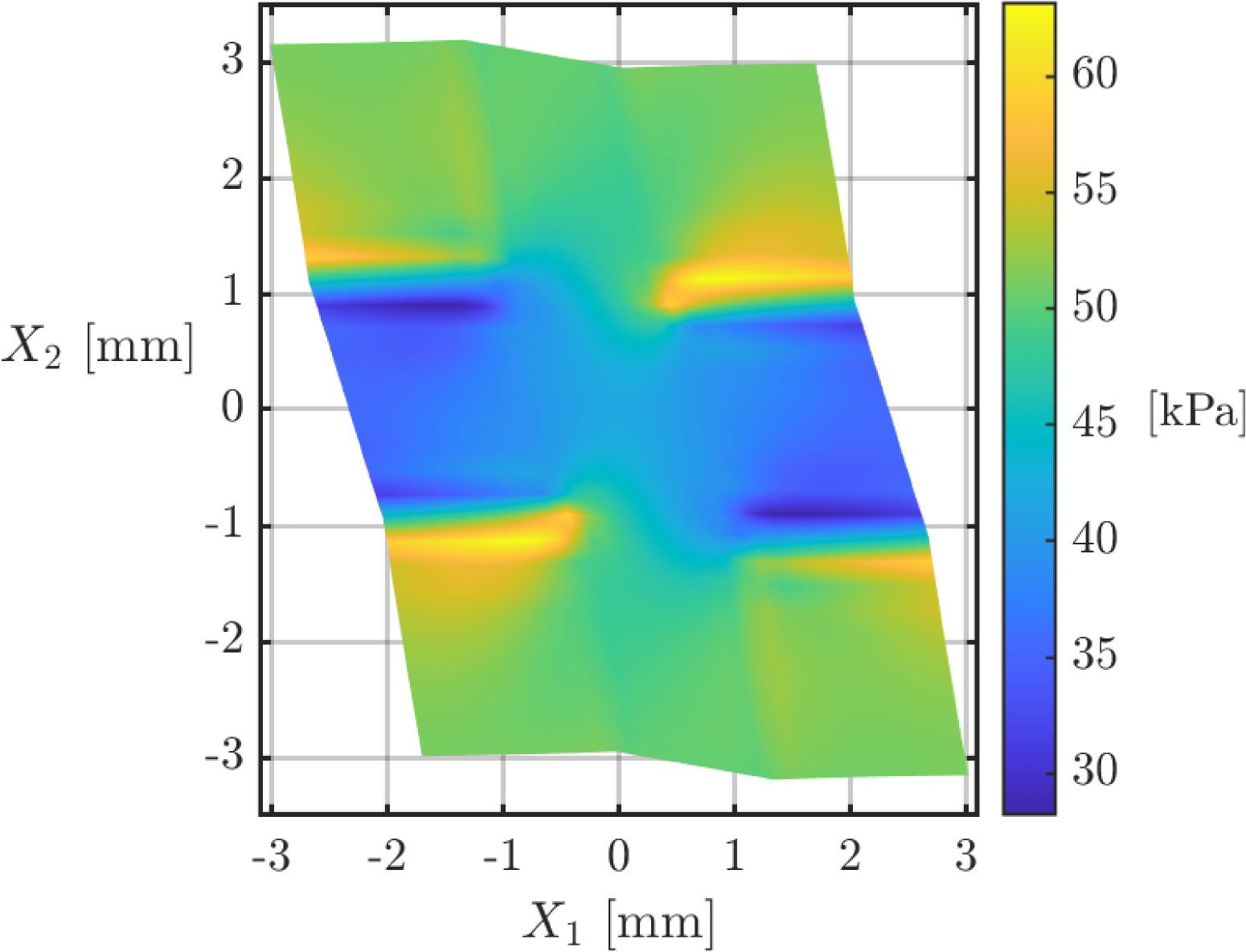} 
\end{tabular}
\end{center}
\caption{Left: von Mises stress distribution at the actual configuration of Cook's membrane. 
Right: von Mises stress distribution actual configuration of an RVE located in the vicinity 
of the upper left corner 
$\vec{P}_4$ of the membrane.}
\label{fig:refCurr}
\end{figure}

To approximate the micro-fluctuation, we apply a classical approach using Dirac pulses for the 
$\vec{X}$-direction according to Eqs.\ \eqref{eq:fluctVaria_dirac}, \eqref{eq:fluctSolution}
and interpolate both $\tilde{\vec{w}}$ in $\tilde{X}$-direction and $\vec{\varphi}$ with 
bilinear four-node elements. 
Thereby, the macro continuum  is resolved with $20\times 20$, while the 
RVE is resolved with $30\times 30$ elements, cf.\ Figure~\ref{fig:refCon}. 
Additionally, we postulate periodic 
conditions on $\partial\RVE$ via a nodal coupling and set 
$\tilde{\vec{w}}=\vec{0}  \;\text{on}\; \mathcal{B}_0\times\delta\Omega_0^{c}$, 
where $\delta\Omega_0^{c}$ consist of the 
four corners $\tilde{\vec{P}}_1=(-3,-3)^\mathrm{T}$, $\tilde{\vec{P}}_2=(3,-3)^\mathrm{T}$, 
$\tilde{\vec{P}}_3=(3,3)^\mathrm{T}$ and $\tilde{\vec{P}}_4=(-3,3)^\mathrm{T}$ of the RVE, to 
restrict rigid body movements%
\footnote{Through periodicity, fixating one corner also fixates opposing 
corners and thus all corners at once.}. 
For the load application, we use an incremental scheme with a total of $n_l\in\{2,12,22,32,42\}$ load steps  
so that in each case a reduced load of 
the form $\vec{T}_k:=k\vec{T}/n_l$ is applied in the $k$th step. The degrees of freedom of the 
respecting results are subsequently used as the starting value for the $(k+1)$th load step to ensure 
the convergence of the Newton iteration. Corresponding results are shown in Figure~\ref{fig:refCurr}.  

In the following, we consider two different settings, wherein the first one we employ the classical 
solution strategy from Section~\ref{sec:classical_sol}. On the micro-level, we utilize the 
termination criterion $\epsilon=10^{-13}$, and on the macro-level, we implement the criterion 
$\epsilon_\mathrm{macro}=10^{-9}$. In the second scenario, we opt for the generalized solution 
strategy outlined in Section~\ref{sec:generalized_sol} and set the termination criterion to 
$\epsilon_\mathrm{macro}=10^{-9}$. A comparison of the number of Newton iterations required in the 
two approaches as a function of the number of load steps is shown in Table~\ref{tab:newtonNumIter}. 
\begin{table}[ht]
\caption{Number of global Newton iterations for different numbers $n_l$ of total load steps.}
\begin{center}
\begin{tiny}
\begin{tabular}[t]{|c|c|c|c|c|c|}
\hline
$n_l$       & 2 & 12 & 22 & 32 & 42 \\
\hline
generalized & 6 &  4 &  4 &  4 &  4 \\
\hline
classical  & /  & /  &  4 &  4 &  4 \\
\hline
\end{tabular}
\end{tiny}
\end{center}
\label{tab:newtonNumIter}
\end{table}
We remark that the number of iterations remains the same from step to step so that the 
values represent the number of iterations in each load step. In addition, the cases in which 
the Newton method was aborted unsuccessfully are indicated by a slash. Thereby, we define 
the procedure as failed if the macro-residual $\|\vec{R}_\mathrm{macro}\|$ exceeds the value $10^{10}$ during 
the iteration. As can be seen, the Newton method terminates from $n_l=22$ using the classical 
solution method, while the generalized method terminates successfully from $n_l=2$. It should be noted that 
the step numbers are minimal in the sense that the scheme does not converge for $n_l=1$ and 
$n_l=21$ using the generalised and the classical solution strategy, respectively. 
Eventually, Table~\ref{tab:newtonIter} shows an example of the convergence 
behaviour of the Newton method for the two solution procedures to give an idea of the processes. 
\begin{table}[!ht]
\caption{Convergence of the Newton method in different load steps for a load application with a 
total of $22$ load steps. The residual is specified in the unit Newton (N). }
\begin{center}
\begin{tiny}
\begin{tabular}[t]{|c|c|c|c|}
\hline
\multirow{3}{*}{$k$} & \multicolumn{3}{c|}{load-step $1$} \\ 
\cline{2-4}
& \multirow{2}{*}{generalized} & \multicolumn{2}{c|}{classical} \\
&  & Macro & RVE \\
\hline
$0$  & $1.54\cdot 10^{1}$   &  $1.54\cdot 10^{1}$    & $2.07\cdot 10^{-15}$ \\
\hline
$1$  & $3.84\cdot 10^{1}$   &  $3.84\cdot 10^{1}$    & $7.00\cdot 10^{-4}$   \\
     &                      &                        & $9.92\cdot 10^{-7}$  \\
     &                      &                        & $1.20\cdot 10^{-12}$  \\
     &                      &                        & $2.34\cdot 10^{-15}$ \\
\hline 
$2$  & $1.92\cdot 10^{-2}$  &  $1.52\cdot 10^{-2}$   & $2.06\cdot 10^{-1}$  \\ 
     &                      &                        & $5.24\cdot 10^{-2}$  \\
     &                      &                        & $7.57\cdot 10^{-3}$ \\
     &                      &                        & $2.71\cdot 10^{-4}$ \\
     &                      &                        & $3.97\cdot 10^{-7}$  \\
     &                      &                        & $9.60\cdot 10^{-13}$ \\
     &                      &                        & $2.25\cdot 10^{-15}$ \\
\hline 
$3$  & $1.06\cdot 10^{-7}$  &  $7.61\cdot 10^{-7}$   & $2.01\cdot 10^{-1}$  \\ 
     &                      &                        & $4.92\cdot 10^{-2}$ \\
     &                      &                        & $6.79\cdot 10^{-3}$ \\
     &                      &                        & $2.14\cdot 10^{-4}$ \\
     &                      &                        & $2.48\cdot 10^{-6}$ \\
     &                      &                        & $3.70\cdot 10^{-13}$ \\
     &                      &                        & $2.34\cdot 10^{-15}$ \\
\hline 
$4$  & $4.61\cdot 10^{-10}$ &  $3.73\cdot 10^{-10}$  & $2.01\cdot 10^{-1}$ \\ 
     &                      &                        & $4.92\cdot 10^{-2}$ \\
     &                      &                        & $6.79\cdot 10^{-3}$ \\
     &                      &                        & $2.14\cdot 10^{-4}$ \\
     &                      &                        & $2.49\cdot 10^{-7}$ \\
     &                      &                        & $3.71\cdot 10^{-13}$ \\
     &                      &                        & $2.40\cdot 10^{-15}$ \\
\hline
\end{tabular}

\hspace{3mm}
\begin{tabular}[t]{|c|c|c|c|}
\hline
\multirow{3}{*}{$k$} & \multicolumn{3}{c|}{load-step $22$} \\ 
\cline{2-4}
& \multirow{2}{*}{generalized} & \multicolumn{2}{c|}{classical} \\
&  & Macro & RVE \\
\hline
$0$  & $1.54\cdot 10^{1}$   &  $1.54\cdot 10^{1}$  & $2.29\cdot 10^{-12}$  \\
\hline
$1$  & $1.17\cdot 10^{1}$   &  $1.17\cdot 10^{1}$   & $5.63\cdot 10^{-4}$  \\ 
     &                      &                       & $6.11\cdot 10^{-7}$  \\
     &                      &                       & $4.38\cdot 10^{-13}$ \\
     &                      &                       & $2.31\cdot 10^{-15}$ \\
\hline
$2$  & $8.97\cdot 10^{-3}$  &  $8.16\cdot 10^{-3}$  & $5.19\cdot 10^{-6}$  \\ 
     &                      &                       & $3.47\cdot 10^{-12}$ \\
     &                      &                       & $2.43\cdot 10^{-15}$ \\
\hline
$3$  & $7.11\cdot 10^{-7}$  &  $5.24\cdot 10^{-7}$  & $2.66\cdot 10^{-7}$  \\ 
     &                      &                       & $1.02\cdot 10^{-13}$ \\
     &                      &                       & $2.65\cdot 10^{-15}$ \\
\hline
$4$  & $4.18\cdot 10^{-10}$ &  $6.00\cdot 10^{-10}$ & $1.75\cdot 10^{-13}$ \\ 
     &                      &                       & $2.36\cdot 10^{-15}$ \\
\hline
\end{tabular}
\end{tiny}
\end{center}
\label{tab:newtonIter}
\end{table}
The value of the macro-residual  
in the $k$th iteration step is given under ``generalized'' and ``classical'', respectively, while 
the iteration progression for the classical staggered scheme at the micro level is given in the column 
``RVE''. The quantities shown depict the course of convergence in relation to an exemplary integration point, 
whereby a point was selected in each case for which the maximum number of iterations was required 
to achieve the termination criterion. 

\section{Conclusions}\label{sec:conclusions}
The variational formulation of the multiscale system leads to a set of partial differential equations in a higher-dimensional space. Even constant shape functions of the fluctuations in direction of the macroscale match the results of Delta Dirac functions evaluated at every Gauss points,  although the number of RVEs is dramatically reduced to one per element. Using higher-order shape functions produces the expected convergence rates, noting that the absolute error for linear shape functions for the fluctuations already surpasses the results of the traditional formulation.  

With regard to \eqref{eq:min1},  we can show that the discrete multiscale problem at hand can be considered as a finite-dimensional and non-convex minimization problem introducing a poly-convex strain energy function on the microscale.  Using this as a discrete objective function,  sophisticated methods like the recursive trust-region multigrid formulations as proposed in Gross \& Krause \cite{krause2009c} can be adapted as well, such that state of the art parallelization techniques can be applied instead of the usual \glqq{}farming\grqq{} on clusters to solve for the physical equilibrium condition on the microscale within every macro Newton step. 

Moreover, it is common to apply a staggered scheme for non-linear multiscale problems, searching first for a physical equilibrium condition on the microscale within each Newton-Raphson step on the macroscale.  However,  as long as the macroscale is not in equilibrium, the physical interpretation of the microscale is questionable. This leads to the curious situation, that a huge amount of additional computational power is required for the microscale solutions, without improving the solution on the macroscale. In the best case, we obtain quadratic convergence on the macroscale, in the worst case, the staggered scheme prevents the solution of the macroscale system e.g.\ for larger external load steps. 

In contrast, the proposed null-space reduction scheme allows for a quadratic convergence simultaneously for both, the micro- and the macroscale without using any additional step in a staggered scheme.  This yields a massive reduction of the computational effort for each iteration.  Additionally, we could demonstrate that the monolithic solution is tremendously more robust. For the Cook's membrane,  the Newton scheme does not converge for the staggered scheme until we apply the external load in 22 load steps, whereas the monolithic scheme requires only two.  Combined with the massive reduction of RVEs to be evaluated using constant shape functions in direction of the macroscale,  the computational power required for the whole simulation is reduced to a fraction compared to the classical formulation.

Moreover, the applied methodology of null-space reduction schemes allows for a rather simple application on all kinds of multiscale problems. The development of a consistent linearization for generalized materials in the context of IGA$^2$, e.g.\ for second or third-order materials is highly tedious, see the Appendix in our previous paper \cite{hesch2022}. The same holds for multiphysical problems.  For the proposed null-space reduction scheme, the construction of a consistent null-space matrix can now be done in a straight-forward manner once the global stiffness matrix of the monolithic macro-micro system has been generated. 

\section*{Acknowledgments}
Support for this research was provided by the Deutsche Forschungsgemeinschaft (DFG), Germany under grant HE5943/26-1. This support is gratefully acknowledged.


	
\bibliographystyle{plain}
\bibliography{literature}

\end{document}